\documentclass{jpp}
\usepackage{graphicx}
\usepackage{natbib}
\usepackage{amsmath}
\usepackage{amssymb}
\usepackage{amsfonts}
\usepackage{amsbsy}
\usepackage{color}

\title{Mechanisms of plasma disruption and runaway electron losses in 
  tokamaks} 
\author[S.S. Abdullaev et al.]
{S. S. Abdullaev$^1$ 
\thanks{Email address for correspondence: s.abdullaev@fz-juelich.de}, \ns
K.H. Finken$^{2}$, K. Wongrach$^2$, M. Tokar$^1$, \ls
 H.R. Koslowski$^1$,  O. Willi$^{2}$, L. Zeng$^3$, and the TEXTOR team}  
\affiliation
{$^1$ Forschungszentrum J\"ulich GmbH,  Institut f\"ur Energie- und
  Klimaforschung - Plasmaphysik, D--52425 J\"ulich, Germany. \\
  $^2$ Institut f\"ur Laser- und Plasmaphysik, Heinrich-Heine 
  Universit{\"a}t D{\"u}sseldorf, Germany, \\ $^3$ Institute of Plasma
  Physics, Chinese Academy of Sciences, 230031 Hefei, China} 
\pubyear{2010}
\volume{650}
\pagerange{119--126}
\date{?; revised ?; accepted ?. - To be entered by editorial office}
\begin{document}

\maketitle

\begin{abstract}
Based on the analysis of data from the numerous dedicated experiments on
plasma disruptions in the TEXTOR tokamak the mechanisms of the formation of
runaway electron beams and their losses are proposed. The plasma disruption is
caused by strong stochastic magnetic field formed due to nonlinearly excited
low-mode number magneto--hydro--dynamics (MHD) modes. It is hypothesized that
the runaway electron beam is formed in the central plasma region confined by an
intact magnetic surface due to the acceleration of electrons by the inductive
toroidal electric field. In the case of plasmas with the safety factor
$q(0)<1$ the most stable runaway electron beams are formed by the intact
magnetic surface located between the magnetic surface $q=1$ and the closest
low--order rational surface $q=m/n>1$ ( $q=5/4$, $q=4/3$, \dots). The thermal
quench time caused by the fast electron transport in a stochastic magnetic
field is calculated using the collisional transport model. The current quench
stage is due to the particle transport in a stochastic magnetic field. The
runaway electron beam current is modeled as a sum of toroidally symmetric part
and a small amplitude helical current with a predominant $m/n=1/1$ component.
The runaway electrons are lost due to two effects: ($i$) by outward drift of
electrons in a toroidal electric field until they touch wall and ($ii$) by the
formation of stochastic layer of runaway electrons at the beam edge. Such a
stochastic layer for high--energy runaway electrons is formed in the presence
of the $m/n=1/1$ MHD mode. It has a mixed topological structure with a
stochastic region open to wall. The effect of external resonant magnetic
perturbations on runaway electron 
loss is discussed. A possible cause of the sudden MHD signals accompanied by
runaway electron bursts is explained by the redistribution of runaway current 
during the resonant interaction of high--energetic electron orbits with the
$m/n=1/1$ MHD mode.        
\end{abstract}
%\pacs{52.25.Fa,05.45.Ac,52.25.Gj,52.55.Dy}

%\tableofcontents
\section{Introduction} 
One of the severe consequences of the plasma disruptions in tokamaks is the
generations of the runaway electron (RE) beams (see, e.g.,
\citet{Wesson_etal_89,Gill_93,Schuller_95,Gill_etal_00,Gill_etal_02,%
Wesson:2004,Boozer_12,Papp_etal_13} and references therein). The REs generated
during the disruptions 
of tokamak plasmas may reach several tens of MeV and may contribute to the
significant part of post--disruption plasma current. The prevention of such RE
beams is of a paramount importance in future tokamaks, especially in the ITER
operation, since it may severely damage a device wall
\citep{Becoulet_etal_13}. \par    

At present there are several proposals to mitigate REs generated during
plasma disruptions. The mitigation of REs by the gas injections has been 
discussed (see, e.g.,
Refs. \citep{ITER:2007_3,Whyte_etal_02,Whyte_etal_03,Bakhtiari_etal_02,%
Bakhtiari_etal_05a,Granetz_etal_07,Bozhenkov_etal_08,Pautasso_etal_09,%
Lehnen_etal_09,Hollmann_etal_10,Reux_etal_10,Lehnen_etal_11}).   
Suppression of REs by the resonant magnetic perturbations (RMPs) has been also
intensively discussed since late 1990s (see, e.g., Refs. 
\citep{Kawano_etal_97b,TokudaYoshino_99,Helander_etal_00,YoshinoTokuda_00,%
Lehnen_etal_08a,Lehnen_etal_09,Hollmann_etal_10,Papp_etal_11a,Papp_etal_12}). 
However, up to now there is no a regular strategy to solve this problem. One
of the reasons is that the physical mechanisms of the formation of REs during
plasma disruptions is still not well--known. The different scenarios of
runaway formation during plasma disruptions are discussed in
literature. Particularly, in Refs.~\citep{Fulop_etal_09,FulopNewton_14} the
possible roles of whistler waves on the generation of REs and Alfv\'enic wave
instabilities driven by REs have been discussed. \par     

There were numerous dedicated experiments to study the problem of runaway
current generation during plasma disruptions triggered by massive gas
injections (MGI) in the TEXTOR tokamak (see, e.g., 
\citep{Forster_etal_12b,Zeng_etal_13,Wongrach_etal_14}), in KSTAR tokamak  
\citep{Chen_etal_13}, the JET tokamak \citep{Plyusnin_etal_06,Lehnen_etal_11},
in DIII-D \citep{Hollmann_etal_10,Commaux_etal_11,Hollmann_etal_13}, Alcator
C-Mod \citep{Olynyk_etal_13}, and others. In these works the dependencies of RE
generation on the toroidal magnetic field, on the magnetic field fluctuations,
on the species of injection gases have been investigated. Particularly, in
KSTAR tokamak \citep{Chen_etal_13} it has been found that there is no the
toroidal magnetic field threshold $B_T<2$ T as was indicated by previous
experiments in other tokamaks. In Ref.~\citep{Izzo_etal_11,Izzo_etal_12} MHD
simulations have been performed to study the confinement REs generated during
rapid disruptions by MGI in DIII-D, Alcator C-Mod, and ITER. Such simulations
with two different MHD codes have been carried out by \citet{Izzo_etal_12} to
analyze shot--to--shot variability of RE currents in DIII-D tokamak
discharges.  \par    

These numerous experiments show the complex nature of plasma disruption
processes especially the formation of RE beams, and its evolution. One of
the important features of this event is its irregularity and variability of RE
beam parameters from one discharge to another one. This indicates the
sensitivity of disruption processes and RE beam formations on initial
conditions which is the characteristic feature of nonlinear processes,
particularly, the deterministic chaotic system. Therefore, {\em ab initio}
numerical simulations of these processes may be not always successful to
understand their mechanisms because of the complexity of computer simulations
of nonlinear processes \citep{Kadanoff_04}. The problems of numerical
simulations of plasma disruptions is comprehensively discussed by
\citet{Boozer_12}.   \par      

In this work we intend to approach to this problem from the point of view of
Hamiltonian chaotic systems, mainly the magnetic stochasticity in a
magnetically confined plasmas \citep{Abdullaev:2014}. Based on the ideas of
these systems and analyses of numerous experimental results, mainly obtained
in the TEXTOR tokamak we propose possible mechanisms of formation and
evolution of RE beams created during plasma disruptions. Since a
self--consistent theoretical treatment of all these processes is very
complicated we developed theoretical models for each stages of a plasma
disruption. These models are used to estimate the characteristic times  of
the thermal and  current quenches, the spatial size of runaway plasma beam
and their decay times, the speed of RE radial drifts, the effect of magnetic
perturbations.    \par  
 
It is believed that the plasma disruption starts due to a large--scale
magnetic stochasticity caused by excited of MHD modes with low poloidal $m$ and
toroidal $n$ numbers, ($m/n= 1/1, 2/1, 3/2$, $5/2, \dots$) (see, e.g.,
\citep{Wesson:2004,Kadomtsev_84,Gill_93,Schuller_95} and references
therein). The heat and particle transports in the strongly chaotic magnetic
field causes the fast temperature drop and ceases the plasma current. However,
at the certain spectrum of magnetic perturbations, for example, at the
sufficiently small amplitude of the $m/n=1/1$ mode the chaotic field lines may
not extend to the central plasma region due to the creation of an intact
magnetic surface. In the case of plasmas with the safety factor $q(0)<1$ at
magnetic axis $\rho=0$ the intact magnetic surface can be located between
magnetic surface $q=1$ and the nearest low--order rational surface $q=5/4$ [or
$q=4/3$, \dots]. This intact magnetic surface confines particles in the
central plasma region and serves as a transport barrier to particles during the
current quench. Electrons in the confined region are accelerated due to large
toroidal electric field and forms the relatively stable of RE beams. \par   

This occurs, for instance, when the plasma disruption initiated by the
heavy Ar gas injection which does not penetrate deep into the plasma,
therefore it does not excite the $m/n= 1/1$ mode with the sufficiently large
amplitude. In contrary, the injection of the lighter noble gases neon and
helium does not generate runaways. The reason is that light gases
penetrate deeper into the plasma and excite the large--amplitude ($m/n=1/1$)
mode. \par  

The existence of an intact magnetic surface and its location depends on the 
radial profile of the safety factor and the spectrum of magnetic
perturbations. The latter sensitively depend on the plasma disruption
conditions and vary unpredictably from one discharge to another during plasma
disruptions. This makes RE formation process unpredictable and may
explain a shot--to--shot variability of the parameters of RE beams.  \par  

The role of the safety factor profile in the formation of RE beams can be
pronounced during disruptions of plasmas with the reversed magnetic
shear. In the plasmas with the non-monotonic radial profiles of the
safety factor it has been observed an improved confinement of energy and
particles due to the internal transport barrier located near the minimal
value of the safety factor, i.e., near the shearless magnetic surface
\citep{Levinton_etal_95,Strait_etal_95}. During disruptions this magnetic
surface acts as a robust magnetic barrier that separates a chaotic magnetic
field formed in outer region from the penetration into the central plasma
region. Electrons confined by the shearless magnetic surface can form a stable
RE beam with a relatively large transversal size. Recently published results
of the disruption experiments in the TFTR tokamak with the reversed magnetic
shear indeed show the formation of a large RE beam with long confinement times
\citep{Fredrickson_etal_15}. \par

Based on this mechanism we study the main three stages of
the post--disruption plasma evolution: the fast thermal quench (TQ), the
current quench (CQ), and the RE beam evolution. The physical
processes during each of these stages will be studied by theoretical
models. These processes are the formation of stochastic magnetic 
field, heat and particle transport in a stochastic magnetic field, the
acceleration of electrons by inductive electric field, the lost mechanisms of
REs, and the effect of internal and external magnetic perturbations.
The short report on this study is to publish in
\citet{Abdullaev_etal_15}.    \par  

The paper consists of eight sections. Mathematical tools and models employed
to study the problems are given in Supplementary part. The numerous data
obtained during the dedicated experiments in the TEXTOR tokamak are analyzed
in Sec.~\ref{description}. Possible mechanisms of plasma disruptions with a RE
beam formation is proposed and analyzed in Sec.~\ref{formation_beam}. The
transport of heat and particles during the fast TQ and the CQ stages of plasma
disruption are studied in Sec.~\ref{transport}. The model of 
a post--disruption plasma beam is proposed in Sec.~\ref{post_disr}. Using this
model a time--evolution of guiding--center (GC) orbits of electrons
accelerating  by the inductive toroidal electric field is studied in
Sec.~\ref{evolution_REs}. Particularly, the change of RE confinement
conditions with decreasing the plasma current and increasing the electron
energy and the outward drift of GC orbits are investigated. The effect of
external and internal magnetic perturbations on the RE confinement are
discussed in Sec.~\ref{Helical}. In the final Sec.~\ref{summary} we give
the summary of obtained results and discuss their consequences.  

\section{Description of plasma disruptions}
\label{description}
 The TEXTOR was a middle size limiter tokamak with the major radius
$R_0=1.75$ m, the minor radius $a=0.46$ cm. The toroidal field $B_0$ can be
varied up to 2.8 T, and the plasma current take up to 600 kA. 
In the experiments the plasma disruptions were triggered in a controlled
way by gas injections using a fast disruption mitigation valve (DMV)
\citep{Bozhenkov_etal_07,Finken_etal_08b,Finken_etal_11,Bozhenkov_etal_11}.
Particularly, the disruptions with REs were triggered by argon (Ar)
injection. The runaway-free disruptions were triggered either by helium or 
neon (He/Ne) injection performed by the smaller valve. The effect of the
externally applied RMPs on the REs generations has been investigated using the
dynamic ergodic divertor (DED) installed in the TEXTOR tokamak. \par 

\begin{figure} 
\centering  %(a) \hspace{5cm} (b) \\
\includegraphics[width=1.0\textwidth]{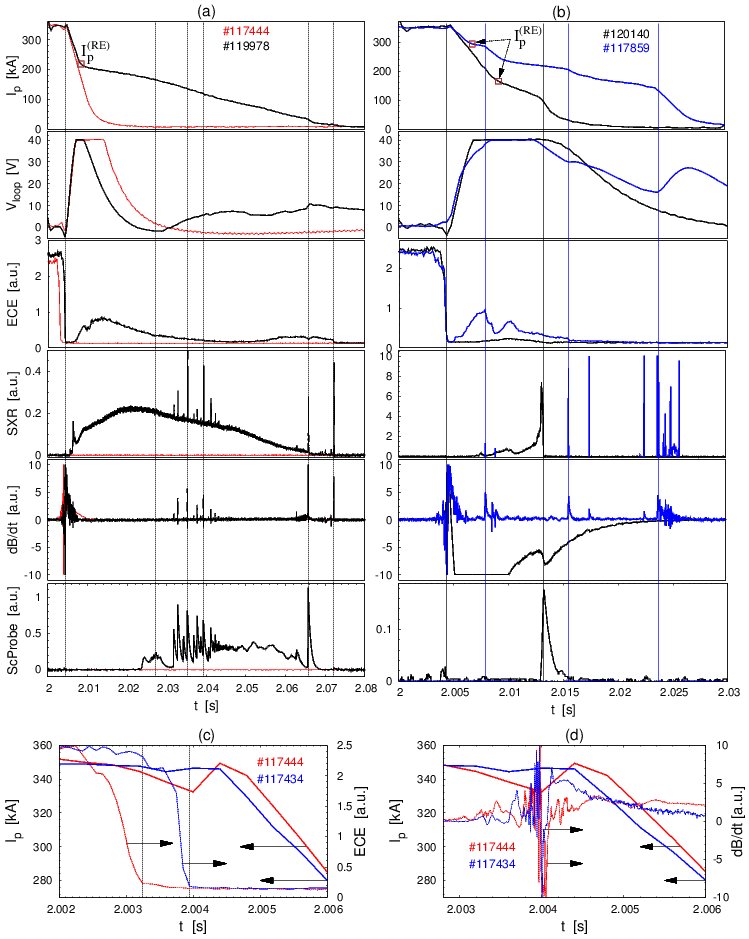} 
\caption{(a) Time evolution of the disruption of the TEXTOR shots \#119978
  (black solid lines) and \#117444 (red curves) (from top to bottom): the
  plasma current, the loop voltage, the ECE signal, the SXR signal, the Mirnov
  signal, and scintillation probe (Sc.Probe) signal. (b) The same but for the
  discharges \#117859 (blue curves)  and \#120140 (black curves). (c)
  Initial stage of the temporal evolution of the plasma current (solid
  curve on the l.h.s. axis), ECE signals (r.h.s. axis); (d) the Mirnov signals
  (r.h.s. axis) during a plasma disruption with (\#117434 and 
  \#117507) and without (\#117444) RE generations.  $I_p^{(RE)}$ is the
  initial value of the plasma current with REs. Disruptions for discharges
  \#117434, \#119978, \#117507, \#117859 \#120140 are initiated by Ar
  injections, and \#117444 by Ne injections.  } 
\label{119978_all}
\end{figure}

Below we analyze the experimental results of discharges with the predisruption
plasma current $I_p=350$ kA and the toroidal field $B_t=2.4$ T.  
Figure~\ref{119978_all} (a) illustrates typical disruptions of the discharges
of the TEXTOR tokamak with and without RE generations. Specifically, it shows
the time evolution of plasma parameters (the loop voltage $V_{loop}$, the
electron cyclotron  emission (ECE), the soft X-ray (SXR) signal, the Mirnov
signal, and scintillation probe signals (ScProbe) during disruptions of the
discharges with REs (\#117434, \#117859, \#119978, \#120140) and without
(\#117444) REs. \par  

There are also some discharges with untypical RE currents and shorter current
decay times. The two examples of such discharges are shown in
Fig.~\ref{119978_all} (b). We will discuss some features of these
discharges at the end of the section. \par   
%--------------------------------Fig.-------------------------------------
%-------------------------------------------------------------------------
The typical behavior of the plasma during the disruptions is following. The
gas (Ar or Ne/He) was injected at the time instant $t=2$ s. One can
distinguish three stages of the disruption with the REs: the first (or fast)
stage in which a sudden temperature drop occurs, in the second stage the
plasma current starts to decay with a higher rate, and in the third stage the
current decay slows down and the current beam with the REs is formed. \par

\subsection{Thermal quench stage}
\label{TQ_stage}
The {\em first fast stage} starts after a few milliseconds (between 2 ms and 5
ms) after the gas injection and ends with a sudden temperature drop (a thermal
quench) in a time interval about one ms as seen from the ECE signals shown in
a detail in Fig.~\ref{119978_all} (c). The Mirnov signals indicating
magnetic activities start just before of this time interval and they last a
few milliseconds until a significant decay of the plasma current for the
RE--free discharges or establishing the current with the REs (see
Fig.~\ref{119978_all} (a)). The close--up views of the ECE signals and the
Mirnov signals in this stage for the discharges with REs (\#117434) and
without REs (\#117444) are shown in Figs.~\ref{119978_all} (c) and (d),
respectively. \par 

For our study it is of importance to analyze in details the difference in 
the time development of the TQ  stage of disruptions without REs and with REs, 
initiated by the massive injection of lighter (Ne) and heavier (Ar) noble
gases.  (i) In Fig.~\ref{119978_all} (c) one can see that the TQ stars roughly 
$2.6~\mu$s, in the former case, and $3.7~\mu$s, in the latter one, after the 
initiation of injection. The ratio of these delay times is of 
$0.7\approx 1/\sqrt{2}$ and can be well explained by the difference in the 
atomic weight $A_g$ of gases in question. Indeed, the flow velocity 
$V_g\sim 1/\sqrt{A_g}$ of neon atoms is by a factor of $\sqrt{2}$ higher than 
that for argon and, thus, atoms enter the plasma after a respectively shorter 
time.  \par 

\begin{figure} 
\centering 
\includegraphics[width=0.60\textwidth]{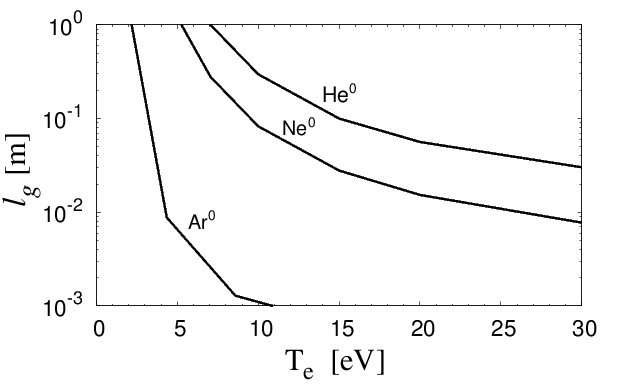}
\caption{Dependence of the penetration length $l_g$ on the electron 
temperature $T_e$ for He, Ne and  Ar atoms. The electron density $n_e=10^{19}$
m$^{-3}$.   }    
\label{penetration}
\end{figure}

(ii) By comparing Figs.~\ref{119978_all} (c) and \ref{119978_all} (d) 
we find that by injection of neon the TQ stage is finished before MHD 
perturbations are triggered. That is in this case TQ is completely due to
cooling induced by the presence of impurity atoms squeezed in a narrow jet and 
penetrating deep enough into the plasma core. The penetration depth of gas 
atoms, $l_g=V_g/\left(k_{ion}n_e\right) $, where $k^0_{ion}$ is the ionization 
rate coefficient. Figure~\ref{penetration} displays $l_g$ versus the electron 
temperature $T_e$ with the electron $n_e=10^{19}$ m$^{-3}$ computed for He, Ne
and  Ar, by using open atomic data base ADAS\footnote{OPEN--ADAS: Atomic Data
and Analysis Structure. http://open.adas.ac.uk/.} for $k^0_{ion}$ and 
assuming that the gas jet has a radial velocity of 2 sound speeds at the room
temperature. One can see that for light gases, He and Ne, $l_g$ can exceed the
minor radius of TEXTOR of $0.46$~m if inside the gas jet the plasma is cooled
down to a temperature of several $eVs$ by energy losses on excitation and
ionization of gas atoms and thermalization of generated electrons, as it is
demonstrated in \citet{KoltunovTokar_11}. The rest of magnetic surfaces is  
cooled down by the heat conduction along magnetic field lines to the jet area 
\citep{TokarKoltunov_13}. Thus, the plasma is cooled down as a whole during a
time of $a/V_g\approx0.5-1$~ms. Only later tearing modes are exited on
numerous resonant magnetic surfaces due to the  growth, as $1/T_e^{3/2}$, of
the plasma resistivity. \par  
%--------------------------------Fig.-------------------------------------
%------------------------------------------------------------------------

(iii) In the case of Ar the penetration depth $l_g$ is much smaller,
mostly due to much larger ionization rate, than for He and Ne atoms and Ar
gas is ionized at the plasma edge. Due to edge cooling very sharp radial
gradients of the plasma resistivity and plasma current density $j$ are
generated. Since the growth rate of tearing modes $\sim \left( dj/d
\rho\right) ^{4/5}$ \citep{Wesson:2004}, MHD modes resonant mostly on outer
magnetic surfaces with the safety factor noticeably large 1 are triggered. The
magnetic field stochastization due to these modes leads to the fast cooling of
the main plasma volume during a time of $0.1-0.2$~ms (see
Sec.~\ref{heat_trans}). This explains why in shot 117434 with Ar injection the 
temperature drop happens although later but faster  than in shot 117444 with
Ne, and MHD activity starts to develop even before the TQ, see
Figs.~\ref{119978_all} (c) and (d). \par  

\subsection{Current quench stage}
The {\em second stage} of the plasma disruption begins with the current decay
within a millisecond after the TQ. Particularly, for the
discharges \#117434 and \#117444 the current decay starts in $0.47\times
10^{-3}$ s and $0.87\times 10^{-3}$ s, respectively, after the temperature
drop (see Figs.~\ref{119978_all} (a) and (c)). In discharges without the RE
formation the current decays with the same rate until it completely disappears
in a few millisecond.  In the discharges with the RE formation the strong
current decay stops at the certain value of $I_p=I_p^{(RE)}$ and replaced by
slower decay. The initial RE current $I_p^{(RE)}$ is shown in
Figs.~\ref{119978_all} (a) and (b). In this stage the loop voltage starts to
rise due to inductive electric field opposing to the current decay. \par

\begin{figure} 
\centering % (a) \hspace{5cm} (b) \\
\includegraphics[width=0.60\textwidth]{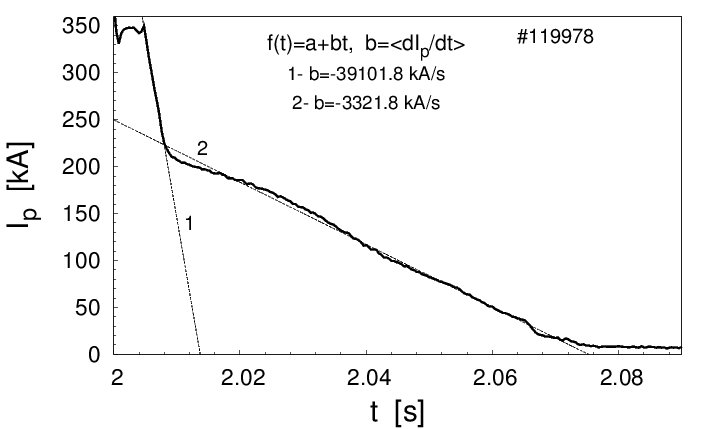}
\caption{(a) Determination of the average current decay rate $\langle dI_p/dt
  \rangle$ at the second and the third stages of the plasma disruptions. Solid
curve describes the measured time--evolution of the current $I_p(t)$, dashed
straight line $f(t)=a-bt$ approximates the average current decay. The
coefficient $b$ gives the estimation of $|\langle dI_p/dt \rangle|$. Curve 1
corresponds to the current decay stage, and curve 2 corresponds to the RE
current decay stage.  }   
\label{Ip_decay}
\end{figure}

The time dependence of the plasma current $I_p$ in this stage for all
discharges is well approximated by the linear function of time
$I_p=I_{p0}-bt$, where the coefficient $b=-\langle dI_p/dt \rangle$ determines
the average current decay rate. The scheme of determination of $b=|\langle
dI_p/dt \rangle|$ is shown in Fig.~\ref{Ip_decay}. 
%--------------------------------Fig.-------------------------------------
%------------------------------------------------------------------------
The values of the current decay rate $\langle dI_p/dt \rangle$ during the
CQ and RE plateau regimes, the initial RE current $I_p^{(RE)}$ for
a number of discharges are listed Table~\ref{Table_Ipt}. It also shows the
time $t_{max}$ when the applied RMPs, i.e., the DED current $I_{ded}$ reaches
its maximal value, and the toroidal mode $n$ of the RMPs. \par 
%--------------------------------Fig.-------------------------------------
\begin{figure} 
\centering  (a) \hspace{6cm} (b) \\
\includegraphics[width=1.0\textwidth]{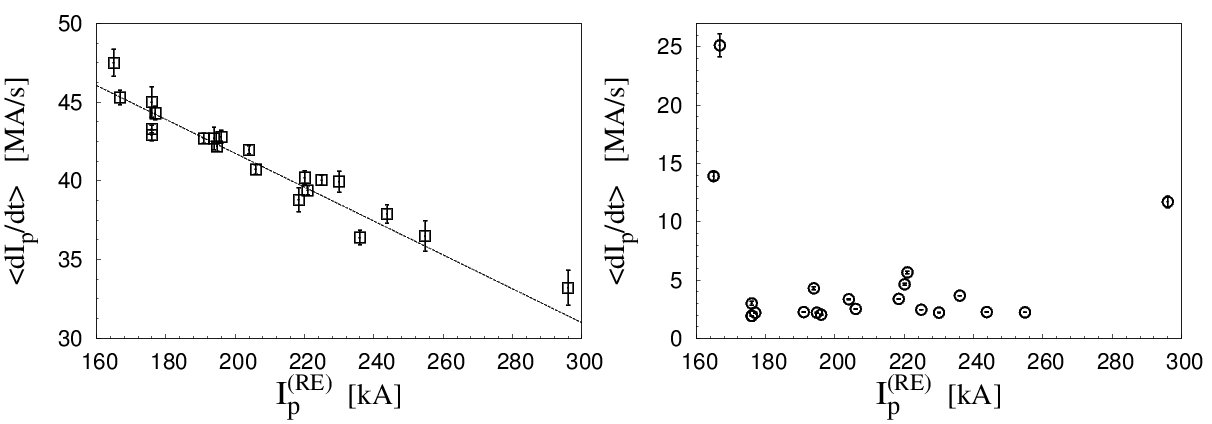}
\caption{(a) Dependence of the average current decay rate $|\langle dI_p/dt
  \rangle|$ on the initial RE current $I_p^{(RE)}$ in the current quench
  regime for a several discharges; (b) The same as in (a) but in the RE
  current decay stage.  }    
\label{Ip_REdecay}
\end{figure}
%------------------------------------------------------------------------
The dependencies of $|\langle dI_p/dt \rangle|$ on the initial RE current
$I_p^{(RE)}$ for a number of discharges are plotted in Fig.~\ref{Ip_REdecay}
(a) and (b) in the CQ regime and the RE current decay stage,
respectively. The current decay rate $|\langle dI_p/dt \rangle|$ for all
discharges are the same order and lies between $32$ and $50$ [MA/s] as listed
in the 2-nd column of Table~\ref{Table_Ipt} and shown in Fig.~\ref{Ip_REdecay}
(a). The highest values of $|\langle dI_p/dt \rangle|$ is observed for the
discharges without REs and with the lowest values of the RE current $I_p^{(RE)}$.
As can see from Fig.~\ref{Ip_REdecay} there is a clear regular dependence of
the current decay rate on the initial RE current, which can be fitted by a
linear function,  $|\langle dI_p/dt \rangle| \approx A-B I_p^{(RE)}$ with
constant parameters $A, B$. In Sec.~\ref{particle_trans} we will discuss the
possible mechanism of such a current decay related with the transport of
particles in a stochastic magnetic field.  \par   
%-------------------------------------------------------------------------
\begin{table} 
\begin{center}
\begin{tabular}{lccccc} 
Discharge No.&~~$2-$nd stage ~~&~~$3-$rd stage & $I_p^{(RE)}$ [kA] & DED
$t_{max}$, $n$, $I_{ded}$ \\[3pt]\hline 
117434 & $40.2\pm 0.45$ & $4.63 \pm 0.09$ & 220.1 & no \\ 
117444 & $47.8\pm 0.21$ & N/A             & N/A   & no \\
117507 & $42.7\pm 0.71$ & $4.29 \pm 0.14$ & 193.9 & no \\
117527 & $39.4\pm 0.29$ & $5.65 \pm 0.12$ & 220.9 & no \\ 
117543 & $50.2\pm 0.19$ & NA              & N/A   & no \\ 
117859 & $33.2 \pm 1.11$& $11.70\pm 0.50$ & 296.0 & no \\
119868 & $40.05\pm 0.31$& $2.45 \pm 0.01$ & 224.9 & no \\ 
119869 & $39.96\pm 0.67$& $2.20 \pm 0.02$ & 230.0 & 2.02 s, $n=1$, 1 kA  \\
119870 & $36.5 \pm 0.96$& $2.23 \pm 0.01$ & 254.8 & 2.02 s, $n=1$, 1.5 kA \\
119874 & $37.9 \pm 0.60$& $2.25 \pm 0.01$ & 243.8 & no \\
119877 & $45.3 \pm 0.47$& $25.13\pm 0.98$ & 166.8 & $1.9$ s, $n=1$, 2 kA \\ 
119978 & $38.8 \pm 0.76$& $3.38 \pm 0.02$ & 218.4 & no \\
119990 & $41.98\pm 0.23$& $3.35 \pm 0.03$ & 204.0 & no \\
120106 & $42.8 \pm 0.42$& $2.03 \pm 0.03$ & 196.0 & $2.0$ s, $n=2$, 4 kA \\
120107 & $40.73\pm 0.29$& $2.52 \pm 0.03$ & 206.0 & $2.0$ s, $n=2$, 4 kA \\
120108 & $42.71\pm 0.32$& $2.25 \pm 0.02$ & 191.0 & $1.9$ s, $n=2$, 4 kA \\
120109 & $42.91\pm 0.35$& $1.93 \pm 0.02$ & 176.0 & $1.9$ s, $n=2$, 4 kA \\
120123 & $36.4 \pm 0.47$& $3.66 \pm 0.01$ & 236.0 & no \\
120126 & $43.3 \pm 0.25$& $1.92 \pm 0.03$ & 176.0 & $2.0$ s, $n=2$, 7 kA \\
120134 & $45.0 \pm 0.97$& $2.99 \pm 0.17$ & 176.0 & $2.0$ s, $n=2$, 7 kA \\ 
120135 & $44.3\pm 0.43$ & $2.20 \pm 0.05$ & 177.0 & $2.0$ s, $n=2$, 7 kA\\
120140 & $47.5\pm 0.86$ & $13.91\pm 0.35$ & 165.0 & $1.97$ s, $n=2$, 6 kA \\
120141 & $42.2\pm 0.30$ & $2.20 \pm 0.03$ & 194.8 & $2.07$ s, $n=2$, 6 kA \\
 \hline 
\end{tabular}
\caption{Parameters of discharges: 1-st column $-$ the discharge number; 2-nd
  and 3-rd columns $-$ the average decay rates  $|\langle dIp/dt \rangle|$ [in
  MA/s] of the 
  plasma current $I_p(t)$ in the second and the third stages; 4-th column $-$
  the initial current of the RE beam $I_p^{(RE)}$; 5-th column shows the
  parameters of the RMPs, a time $t_{max}$ when the DED current reaches its
  maximum values $I_{ded}$, the toroidal mode $n$. Note, that the 
  discharges \#117444 and \#117543 are RE-free. }  
\label{Table_Ipt}
\end{center}
\end{table}
%-------------------------------------------------------------------------
\subsection{RE plateau stage}
In the {\em third stage} (RE plateau) of the disruption the rapid current
decay is replaced by it's slow decay and it starts the formation of the REs
due to the acceleration of electrons in the inductive toroidal electric field
and the secondary generation of REs. The values of the current decay rate
$|\langle dI_p/dt \rangle|$ along with the initial values of the plasma
current $I_p^{(RE)}$ in  this stage for several discharges are listed in the
3--rd and the 4--th columns of Table~\ref{Table_Ipt} and plotted in
Fig.~\ref{Ip_REdecay} (b). The average values of $|\langle dI_p/dt \rangle|$
for almost all discharges are confined in the interval (2.2, 5.6) MA/s, i.e.,
in one order lower than the current decay rate in the second stage. The values
of $I_p^{(RE)}$ are also confined in the range between 170 kA and 260 kA,
except of some untypical discharges, which will be discussed in the next
subsection.  These values of $|\langle dI_p/dt \rangle|$ and $I_p^{(RE)}$ 
are close to the ones observed in the similar experiments in the DIII-D
tokamak (see, e.g., \citep{Hollmann_etal_10,Hollmann_etal_13}). \par 

One should also note that in the RE plateau stage at certain time instants
one observes also a sudden current drop accompanied by magnetic
activity and RE bursts as seen from Figs.~\ref{119978_all} (a) and (b) (see
also e.g., Refs.~\citep{Gill_etal_00,Forster_etal_12b}). These events are 
probably related to the nonlinear interaction of high--energetic electrons
with MHD modes which leads formation a stochastic layer at the beam edge open
to the wall. We will discuss this phenomenon in Sec.~\ref{MHD_mode_gen}.
In the {\em final termination stage} one observes the quick RE current losses
accompanied by magnetic activity. \par 

\subsection{Untypical discharges with REs}  
As was mentioned above there are several untypical discharges for which the
rates $\langle{dI_p/dt}\rangle$ take highest or lowest values (see
Figs.~\ref{119978_all}~(b), Fig.~\ref{Ip_REdecay} (b), and Table
\ref{Table_Ipt}). Particularly, the current decay rate (in the 2-nd 
stage) for \#117859 is lowest and highest for discharges \#119877,
\#120140. The RE current decay rate (in the 3-rd stage) for these discharges
takes highest values. The quantity $I_p^{(RE)}$ takes the lowest value for the
discharges \#119877, \#120140 and the highest value for \#117859 as shown in
Fig.~\ref{Ip_REdecay} (b). One can notice strong spikes in the SXR signals of
these discharges in compared to typical discharges (see Fig.~\ref{119978_all}
(a) and (b)). Moreover, the above mentioned bursts of REs accompanied by
magnetic activities are more pronounced in these discharges. We will discuss
the peculiarity of these discharges in  Secs.~\ref{formation_beam} and
\ref{post_disr}. \par         
  
\subsection{Effect of the RMPs on RE generation}
In a number of discharges the effect of the DED of the TEXTOR (see Sec. 5.2 of
Supplementary Part) on the RE generation has been studied. It was found that
the RMPs do not completely eliminate the RE formation, but it can increase the
decay rate $\langle{dI_p/dt}\rangle$ and decrease $I_p^{(RE)}$. This effect 
depends on the operational mode $n$, the amplitude of the DED current
$I_{DED}$, on the time $t_{max}$ when the maximal DED current is reached. As
seen from Table~\ref{Table_Ipt} the maximal effect is obtained when the
maximal $I_{ded}$ is reached before the gas injection at $t=2.0$ s, i.e.,
$t_{max} \leq t=2.0$~s. However, at $t_{max} > t=2.0$~s the RMPs does not
affect at al or it is a very weak. Other experimental observations in the
TEXTOR also confirm these observations \citep{Koslowski_etal_14}. We will
discuss this problem in Sec.~\ref{Effect_DED}. \par    

\section{Formation of a confined plasma beam}
\label{formation_beam}

\subsection{Main conjecture}
It is believed that the plasma disruption is caused by a large scale magnetic
stochasticity of field lines due to interactions of nonlinearly destabilized 
MHD modes \citep{Carreras_etal_80,Kadomtsev_84,Lichtenberg_84,
Fukuyama_etal_92,Wesson:2004,Kruger_etal_05,White:2014}). The global
stochasticity are mainly due to the interactions of coupled MHD modes with low
($m,n$) numbers: ($m=1,2,\dots$), ($n=1,2 \dots$). The structure of a
stochastic magnetic field mainly depends of the amplitudes $B_{mn}$ of MHD
modes and the radial profile of the safety factor $q(\rho)$, where $\rho$ is
the minor radius of a magnetic surface. Depending on these parameters the
stochastic magnetic field may fill entirely plasma region so that the plasma
particles are transported out along chaotic magnetic field lines which leads
to cease of a plasma current. However, at certain conditions the stochastic
magnetic field may not extend up to the central plasma region due to the
formation of a magnetic barrier by the outermost intact magnetic surface at
$\rho_c$.  The electrons confined by this magnetic surface are accelerated by
the toroidal electric field induced the current decay from outer plasma
region, and thus forming a RE beam. Let $I_p(\rho)$ be the plasma current
flowing inside the magnetic surface of radius $\rho$,    
%-------------------------------------------------------------------------
\begin{align}  \label{Ip_model}
I_p(\rho)=2\pi \int_0^{\rho} j(\rho) \rho d\rho, 
\end{align}
%-------------------------------------------------------------------------
where $j(\rho)$ is the current density. Then the initial RE current
$I_p^{(RE)}$ is mainly determined by the predisruption 
plasma current distribution $I_p(\rho)$ confined by the intact magnetic
surface $\rho_c$, i.e., $I_p^{(RE)} \approx I_p(\rho_c)$.\par

As will be shown in Secs.~\ref{evolution_REs} and \ref{Helical} the decay
of the RE beam mainly depends on the two effects: the outward drift of RE
orbits induced by the toroidal electric field $E_\varphi$ and their resonant
interactions with helical magnetic perturbations. The outward drift velocity
$v_{dr}$ is determined of $E_\varphi$ and the RE current, 
%-------------------------------------------------------------------------
\begin{align}  \label{v_dr_RE}
v_{dr} \propto E_\varphi/I_p^{(RE)} \propto E_\varphi/\rho_c^2.
\end{align}
%-------------------------------------------------------------------------
The most stable RE beams are expected to form when the corresponding drift
velocity is lowest and the low--order rational magnetic surfaces within the RE
beam are absent or only one.   \par   

\subsection{Possible generic structures of stochastic magnetic fields}

Below we study possible structures of the stochastic magnetic field which may
lead to the formation of the RE beams. We consider the two type of the safety
factor profiles of $q(\rho)$: ($i$) the monotonic radial profile and ($ii$) the
non--monotonic radial profile, corresponding to the plasmas with the reversed
magnetic shear. \par 

The models for the radial profiles of the plasma current $I_p(\rho)$, the
safety factor $q(\rho)$ of the pre-disruption equilibrium plasma, and the MHD
magnetic perturbations are given in Sec.~3 of Supplementary part. The
perturbation magnetic field simulating low--mode number MHD modes is given by
the toroidal component of the vector potential 
%---------------------------------Eq.()---------------------------------
\begin{align}  \label{A_MHD_sum}
&A_\varphi^{(1)}(R,Z,\varphi,t) = - \frac{R_0^2}{R} \sum_{mn} m^{-1}
a_{mn}(\rho)  \cos\left(m\vartheta -n \varphi + \Omega_{mn}t \right), \cr 
&a_{mn}(\rho)= B_{mn} U_{mn}(\rho),
\end{align}
%----------------------------------------------------------------------------
with the mode amplitudes $B_{mn}$ and rotation frequencies $\Omega_{mn}$. Here
$B_0$ is toroidal field strength, $R_0$ is the major radius $R_0$, and the
functions $U_{mn}(\rho)$ describes the radial profiles of modes. \par 

One should note that the structure of magnetic field lines in the presence of
magnetic perturbations is less sensitive to the radial profiles of
$U_{mn}(\rho)$. It is mainly determined by the safety factor profiles and the
mode amplitudes $a_{mn}(\rho)$ at the resonant surfaces $\rho=\rho_{mn}$,
$q(\rho_{mn})=m/n$ (see Sec. 7.2 of Supplementary part).  \par 

\textbf{Monotonic radial profile of  $q(\rho)$: The case  $q(0)<1$.}
The typical TEXTOR plasma has the monotonic safety factor profile with the
value $q(0)<1$ at the magnetic axis $\rho=0$. In this plasma, the $m/n=1/1$
mode should play an important role on the structure of stochastic field lines
near the plasma center. At low amplitudes of this mode the global stochastic
field lines may not reach $q=1$ magnetic surface and may form a confined region
about the plasma center where REs can be generated. At high amplitudes of the
$m/n=1/1$ mode the stochastic field lines may cover entire plasma region with
no confined particles.  \par  

As was mentioned above in TEXTOR experiments plasma disruptions with REs were
deliberately caused by the injection of Ar gas while the RE--free disruptions
are triggered by He/Ne injection. Experiments show that the penetration
lengths of atoms depends on their atomic weights \citep{Bozhenkov_etal_08}: He
(or Ne) atoms penetrate deeper into plasma than Argon atoms. The injection of
these gases may finally give rise to different spectra of amplitudes of MHD 
modes. One can expect that the amplitude of the $m/n=1/1$ MHD mode excited by
the He/Ne injection is higher than in the case of Argon gas injection. \par  

\begin{figure} 
\centering  (a) \hspace{6cm} (b) \\ 
\includegraphics[width=0.99\textwidth]{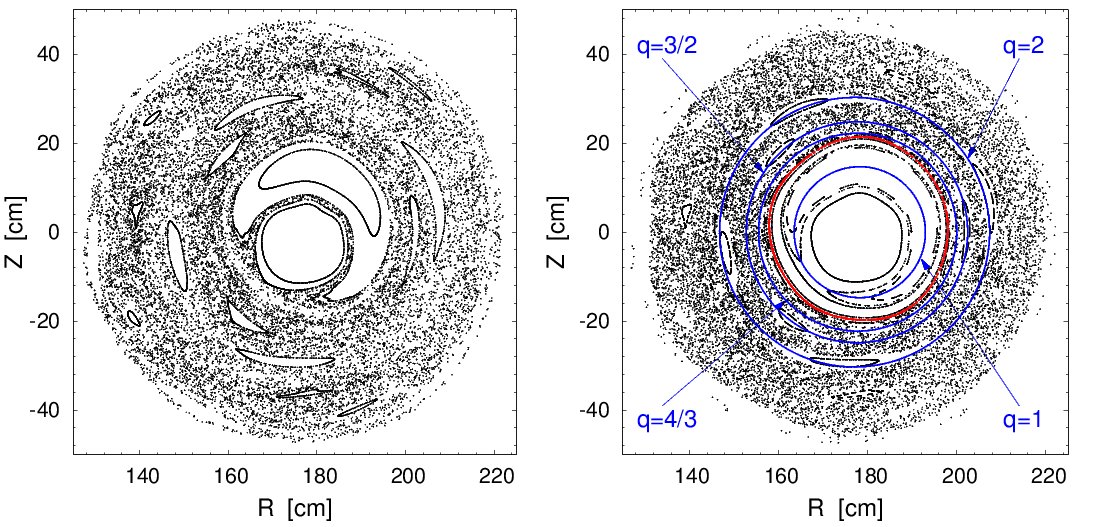}  
\caption{Poincar\'e sections of magnetic field lines in a pre--disruption
  plasma caused by several MHD modes: (a) the mode amplitudes
 $b_{mn}=B_{mn}/B_0$ are $(1,1,1,1)\times \epsilon_{MHD}$; (b) $b_{mn}=B_{mn}/B_0$ are
  $(0.5,1,1,1) \epsilon_{MHD}$. Red curve corresponds to the last intact
  magnetic surface, blue curves are the resonant magnetic surfaces $q=1$,
  $q=4/3$, $q=3/2$, and $q=2$, respectively. The dimensionless perturbation
  parameter $\epsilon_{MHD}=1.5\times 10^{-4}$. The plasma current $I_p=0.35$
  MA, the toroidal field $B_0=2.5$ T, the safety factor at the magnetic axis is
  $q(0)=0.8$. }   
\label{disrup_plasma_sect1}
\end{figure}

The two possible distinct generic structures of a stochastic magnetic field
before the current quench with the RE-free discharge and with the RE discharge
are shown in Figs.~\ref{disrup_plasma_sect1} (a) and (b) by the Poincar\'e
sections of magnetic field lines. It is assumed that the perturbation field
contains several MHD modes: ($m/n=1/1$), ($m/n=2/1$), ($m/n=3/2$), and
($m/n=5/2$). In the case shown in Fig.~\ref{disrup_plasma_sect1} (a) the
normalized mode amplitudes $b_{mn}=B_{mn}/B_0$ are $(1,1,1,1) \times
\epsilon_{MHD}$, and in Fig.~\ref{disrup_plasma_sect1} (b): $(1/2,1,1,1)
\times \epsilon_{MHD}$. The toroidal field magnitude is $B_0=2.5$ T and the
dimensionless perturbation parameter $\epsilon_{MHD}=10^{-4}$. As seen from  
Fig.~\ref{disrup_plasma_sect1} (a) for the larger amplitude of the ($m/n=1/1$)
mode the stochastic magnetic field extends up to the central plasma region 
destroying the separatrix of the $m=n=1$ island. For the low--amplitude of the
($m/n=1/1$) mode shown in Fig.~\ref{disrup_plasma_sect1} (b) the stochastic
magnetic field does not reach the $q=1$ magnetic surface and covers the region
outer the $q=1$ magnetic surface. The last intact magnetic surface $\rho_c$
(red curve) is located between the resonant surfaces $q(\rho_1)=1$ and
$q(\rho_3)=4/3$ (blue curves).  \par       
%--------------------------------Fig.-------------------------------------
%-------------------------------------------------------------------------
As seen from Fig.~\ref{disrup_plasma_sect1} particles in the plasma core are
confined by intact magnetic surfaces located between resonant  surfaces
$q=4/3$ and $q=1$. Plasma beam confined in this area is relatively stable. It 
contains only the $m/n=1/1$ MHD mode which does not lead to a global
stochasticity. The radial transport of particles from the confined area can
take place only due to small--scale turbulent fluctuations and therefore it
has much smaller rate than those in the stochastic zone. The confinement time
of these electrons is sufficiently long enough to be accelerated by the
inductive electric field, thus creating a RE beam. The modeling of the current
of this confined plasma will be discussed in Sec.~\ref{post_disr}.  \par 

\begin{figure} 
\centering  (a) \hspace{6cm} (b) \\ 
\includegraphics[width=0.99\textwidth]{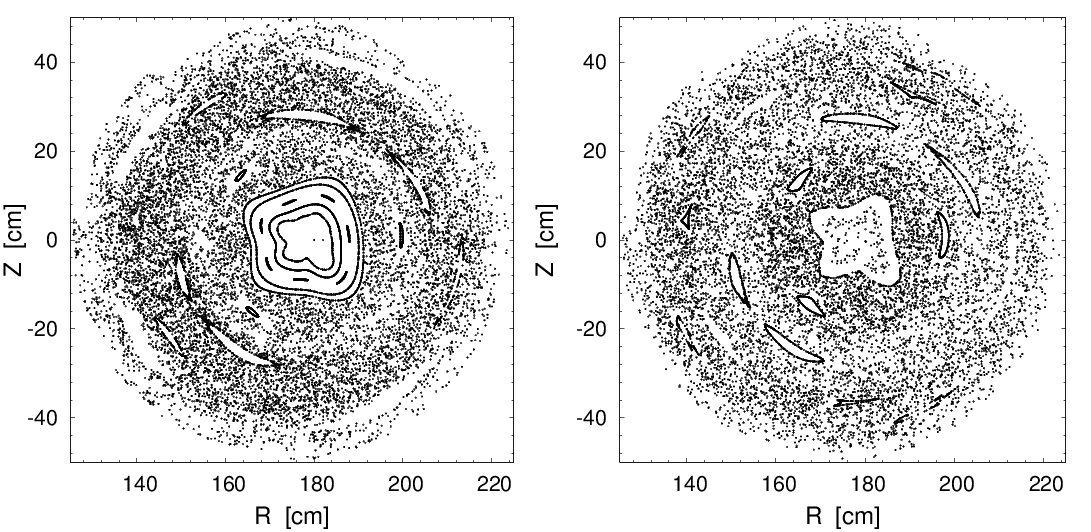}  
\caption{The same as in Fig.~\ref{disrup_plasma_sect1} but for the case
  $q(0)>1$: (a) $q(0)=1.1$; (b) $q(0)=1.2$. It is assumed the magnetic
  perturbation contains the $m/n$ modes ($n=1-3$, $m=1-8$) with the equal
  amplitudes. The dimensionless perturbation parameter
  $\epsilon_{MHD}=1.0\times 10^{-4}$. The plasma parameters are the same as in
  Fig.~\ref{disrup_plasma_sect1}. }    
\label{disrup_plasma_q12}
\end{figure}

\textbf{Monotonic radial profile of  $q(\rho)$: The case  $q(0)>1$.} 
In this case the $m/n=1/1$ mode does not play significant role in the
formation of the stochastic zone in the plasma center. However, the $m/n$ modes
with $n\geq 3$ contribute much to the growth of stochastic zone and shrinkage
of the intact magnetic surface $\rho_c$. The examples of such stochastic
magnetic fields are shown in Fig.~\ref{disrup_plasma_q12} (a) and (b)
corresponding to the values $q(0)=1.1$ and $q(0)=1.2$, respectively.  \par 
%--------------------------------Fig.-------------------------------------
%-------------------------------------------------------------------------
The outward drift velocity $v_{dr}$ of such RE beams is significantly larger
than the one in the case $q(0)<1$. This is because of the smaller RE beam
radius $\rho_c$ and the higher toroidal electric field $E_\varphi$. Such RE
beams decay in shorter times. \par    

\begin{figure} 
\centering  (a) \hspace{6cm} (b) \\ 
\includegraphics[width=0.99\textwidth]{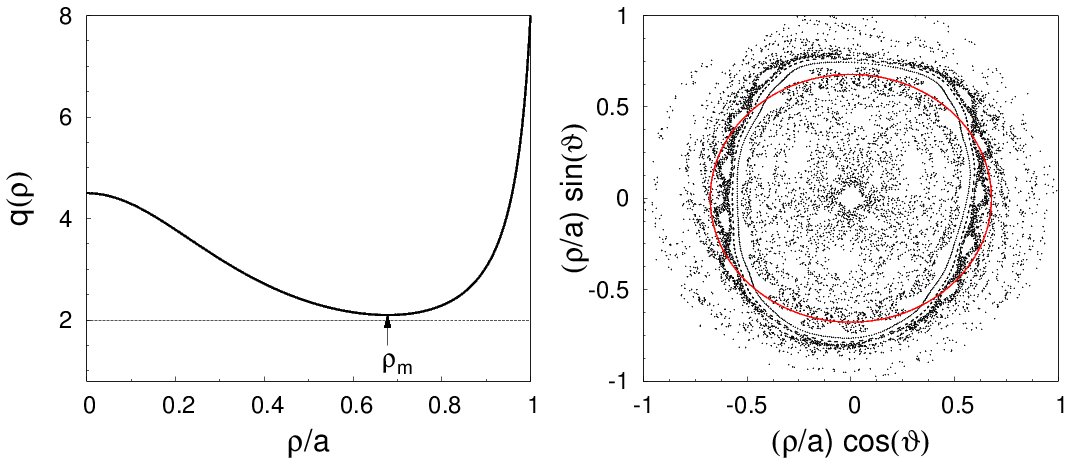}  
\caption{(a) Radial profiles of the safety factor $q(\rho)$ in the plasma with
  the reversed magnetic shear. (b) Poincar\'e sections of magnetic field lines
  in a pre--disruption plasma caused by several MHD modes. Red curve
  corresponds to the shearless magnetic surface. }   
\label{disrup_plasma_rev}
\end{figure}

\textbf{Plasma with reversed magnetic shear}. In this case the safety factor
$q(\rho)$ has a minimal value located at normalized  radius $\rho_m/a\sim$
0.4-0.6 and increases towards the center and the plasma edge.
Figure~\ref{disrup_plasma_rev} (a) and (b) show the example of the
non--monotonic radial profile of the safety factor $q(\rho)$ and the
corresponding Poincar\'e section of stochastic magnetic field lines
(More detailed description of this case is given Sec.~7.3 of the Supplementary
Part). The intact magnetic surface located near the shearless magnetic surface
(red curve), i.e. the magnetic surface with a minimal value of the safety
factor $q(\rho)$, is not broken even at the relatively large magnetic
perturbations. And it confines electrons in the central plasma region. \par
 
Due to the relatively large confined area the RE beam would carry a large
current $I_p^{(RE)}$. According to (\ref{v_dr_RE}) the decay rate of this RE
beam owing to the outward drift would be small. This effect probably explains
the large RE current with a long lifetime observed in the TFTR tokamak during
the disruption of plasmas with the reversed magnetic shear 
\citep{Fredrickson_etal_15}.  \par 
%--------------------------------Fig.-------------------------------------
%-------------------------------------------------------------------------

\subsection{Experimental evidences}
\label{Experimental_evidences}
\begin{figure} 
\centering  
\includegraphics[width=0.8\textwidth]{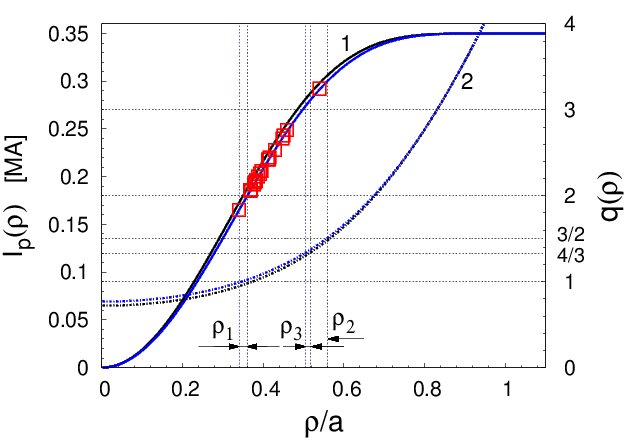} 
\caption{Radial profile of the plasma current $I_p(\rho)$ (\ref{Ip_model})
  (solid curves 1 on l.h.s. axis) and the corresponding safety factor profile
  $q(\rho)$ (dotted curves 2 on r.h.s. axis). The rectangular (red)
  dots correspond to the experimentally measured values of $I_p^{(RE)}$ for
  several TEXTOR discharges. The plasma parameters are $I_p=350$ kA,
  $B_0=2.4$ T, $R_0=1.75$ m, $a=0.46$ m. The values of $q_0=q(0)$ are 0.75 and
  0.8, respectively. The radii $\rho_1$, $\rho_2$, and $\rho_3$ are the
  positions of the rational magnetic surfaces $q(\rho_1)$=1, $q(\rho_2)$=3/2,
  and $q(\rho_3)$=4/3, respectively. }    
\label{disrup_plasma}
\end{figure}

\textbf{Existence of the finite interval of the RE currents $I_p^{(RE)}$}. 
It follows from the conjecture above that the RE current $I_p^{(RE)}$ is mainly
determined by the current distribution $I_p(\rho)$ in the predisruption plasma
confined by the intact magnetic surface $\rho_c$, i.e., $I_p^{(RE)} \approx
I_p(\rho_c)$. Since $\rho_c$ is located between the magnetic surfaces $\rho_1$
and $\rho_3$ corresponding to $q(\rho_1)=1$ and $q(\rho_3)=4/3$, the RE
current $I_p^{(RE)}$ should be in a finite range. This expectation is in
line with the experimental data presented in Fig.~\ref{Ip_REdecay}. One can see
that the range of stable $I_p^{(RE)}$ values  shown in Fig.~\ref{Ip_REdecay} b 
corresponds well to the space between resonant magnetic surfaces with 
$q(\rho_1)=1$ and $q(\rho_3)=4/3$ [or $q(\rho_2)=3/2$]. In 
Fig.~\ref{disrup_plasma} the radial profile of the pre--disruption plasma 
current $I_p(\rho)$ and the corresponding safety factor profile $q(\rho)$ are 
plotted. Also values of $I_p^{(RE)}$ found in other experiments on TEXTOR, see 
Ref.~\citep{Zeng_etal_13}, lie in the same range. \par  
%--------------------------------Fig.-------------------------------------
%-------------------------------------------------------------------------

Since the $q$-value the plasma axis $\rho=0$ is one of the major causes for 
uncertainties in the $q\left( \rho\right) $-profile, in 
Fig.~\ref{disrup_plasma} we show $I$ and $q$ profiles for $q(0)=0.75$ and 
$q(0)=0.8$. These are in the range of $q(0)$ values experimentally measured 
between sawtooth crashes in the TEXTOR tokamak 
\citep{SoltwischStodiek_87,Soltwisch_etal_87a} (see also\citep{Wesson:2004}, 
page 372). The values of $q(0)$ measured after pellet injection in the DIII-D 
tokamak experiments are also close to these values \citep{Izzo_etal_12}. Thus, 
small changes in $q(0)$ still keeps the RE currents
$I_p^{(RE)}$ in the interval $\rho_1< \rho < \rho_3$. The highest and lowest
values of $I_p^{(RE)}$ shown in Fig.~\ref{disrup_plasma} corresponding to the 
discharges \#117859 and \#120140, respectively, lie at the border of region 
$\rho_1< \rho < \rho_3, \rho_2$. They have the shortest duration time for
the RE current decay (see Table~\ref{Table_Ipt} and Figs.~\ref{119978_all}~(b),
\ref{Ip_REdecay}). The presence of a several low--order $m/n=4/3$, $m/n=3/2$,
and $m/n=1/1$ resonant magnetic surfaces within the RE beam may lead to
excitations of the corresponding MHD modes. The interactions of these modes
may lead to the quick loss of REs due to the formation of stochastic zone at
the edge of the RE beam (see Sec.~\ref{Helical}).  \par   

\textbf{Dependence on the level of magnetic perturbations}.
The existence of the intact magnetic surface $\rho_c$ between the $q=1$ and
$q=5/4$ (or $q=4/3$) rational magnetic surfaces and its location depends on
the level magnetic perturbation $\epsilon_{MHD}$ (more exactly on the spectrum
$B_{mn}$). With increase of $\epsilon_{MHD}$ the radius $\rho_c$ shrinks and
it can be broken at the certain critical perturbation level 
$\epsilon_{cr}$. It leads to the total destruction of the confinement of 
plasma particles. This is in agreement with experimental observations on the 
existence of a critical magnetic perturbation level above which the runaway 
beams are not formed \citep{Zeng_etal_13}. \par 

The shrinkage of $\rho_c$ with increasing the magnetic perturbation
$\epsilon_{MHD}$ leads to the decrease of the RE current $I_p^{(RE)}$ since
$I_p^{(RE)} \approx I_p(\rho_c)$. On the other hand if one assumes that the plasma
current decay is caused by the radial transport of particles in the stochastic
magnetic field, its decay rate $dI_p/dt$ should be proportional to the square
of the magnetic perturbation level $\epsilon_{MHD}$, $|\langle dI_p/dt \rangle|
\propto \left|\epsilon_{MHD}\right|^2$ (see
Sec.~\ref{particle_trans}). Therefore, one expects that to the higher 
values of $|\langle dI_p/dt \rangle|$ correspond the lower values of the RE
current $I_p^{(RE)}$. This expectation is in agreement with the experimental
values of these quantities presented in Fig.~\ref{Ip_REdecay} (a). \par 

\begin{figure} 
\centering  
\includegraphics[width=0.6\textwidth]{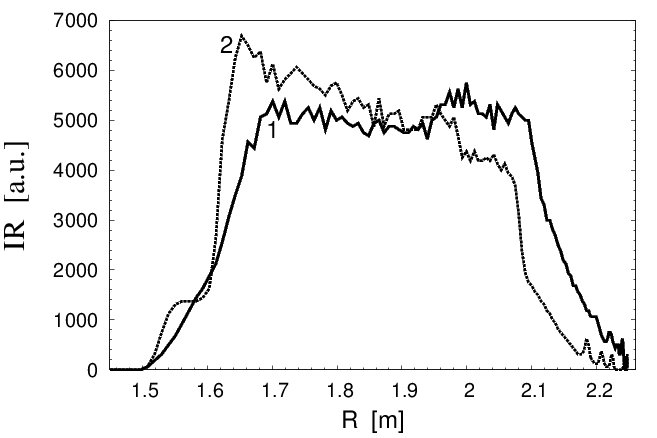} 
\caption{Radial profiles of the synchrotron radiation at the equatorial plane
  $z=0$: curve 1 corresponds to the discharge \#117507 at the time instant
  $t=2.034324$ s and curve 2 $-$ to \#120134 at $t=2.050284$ s.  }   
\label{IR_profile}
\end{figure}

\textbf{Synchrotron radiation pattern}. 
The formation of the RE beam inside the intact magnetic surface can be also
confirmed by the spatial profiles of the synchrotron radiation of high--energy
REs with energies exceeding 25 MeV. Figure~\ref{IR_profile} shows the radial
profiles of infrared radiation of the REs at the equatorial plane $z=0$ for
the two TEXTOR discharges. One can see that radiation is localized inside
finite radial extent corresponding to the central region of plasma within
the $q=4/3$ magnetic surface (see Fig.~\ref{disrup_plasma_sect1} (b)). One
should note the radiation from the plasma edge regions, $1.5\lesssim R
\lesssim 1.6$ and $2.1\lesssim R \lesssim 2.2$ is due to thermal radiation of
the wall elements. The outward shift of the radiation pattern is 
explained to some extend by the drift of RE beams discussed in 
Sec.~\ref{Outward_drift_RE_orbits}.    \par    
%--------------------------------------------------------------------------
%---------------------------------------------------------------------------

Another indication of the formation of confined plasma beam is the rise of the
temperature at the initial stage of the beam formation as seen in the ECE
signals shown in Figs.~\ref{119978_all} (a) and(b). It may occur due to the
Ohmic heating of confined plasma by the induced toroidal electric field or by
superthermal emission from high energy electrons. As thermal electrons are 
converted into runaway ones the beam temperature goes down. \par  

\section{Thermal and current quench stages}
\label{transport}
Our analysis in section 2.1 reveals that both the processes induced
directly by injected atoms and strong radial transport along stochastic
magnetic field lines created by MHD-perturbations can lead to heat losses from
the plasma in the TQ stage of disruption. During this stage the temperature
drops on a timescale of several hundreds of microseconds; current decay time
is of $(4 \div 6)$ ms in RE--free discharges and increases up to 0.1 s in
discharges  with RE generations. \par  

As one can see in Fig.1 the CQ stage before the formation of RE beams 
coincides well with the time interval where magnetic perturbations are 
significant and the particle transport in stochastic magnetic field leads to
the current decay. To study these processes we use the models for stochastic 
magnetic field and collisional transport of test particle described in Sec.~7
of Supplementary part.   \par 

Figures~\ref{disrup_plasma_sect2} (a) and (b) shows the typical Poincar\'e 
sections of field lines of this model in the runaway-free disruption case (a)
and the case with RE generation (b). The perturbation amplitudes
$\epsilon_{mn}$ of all MHD modes, except $(m=1,n=1)$ mode, correspond to the
twice larger value of $\epsilon_{MHD}$ than the case shown in
Fig.~\ref{disrup_plasma_sect1}. For the $(m=1,n=1)$ mode $\epsilon_{mn}$
corresponds to the same value of $\epsilon_{MHD}$. The relation between 
$\epsilon_{mn}$ and $\epsilon_{MHD}$ is $\epsilon_{mn}=
\epsilon_{MHD}b_{mn}/\Psi_a$, where $\Psi_a$ is the toroidal magnetic flux at
the plasma edge (see Sec.~7 of Supplementary part).    \par   
%--------------------------------------------------------------------------
\begin{figure} 
\centering  (a) \hspace{5cm} (b) \\ 
\includegraphics[width=0.99\textwidth]{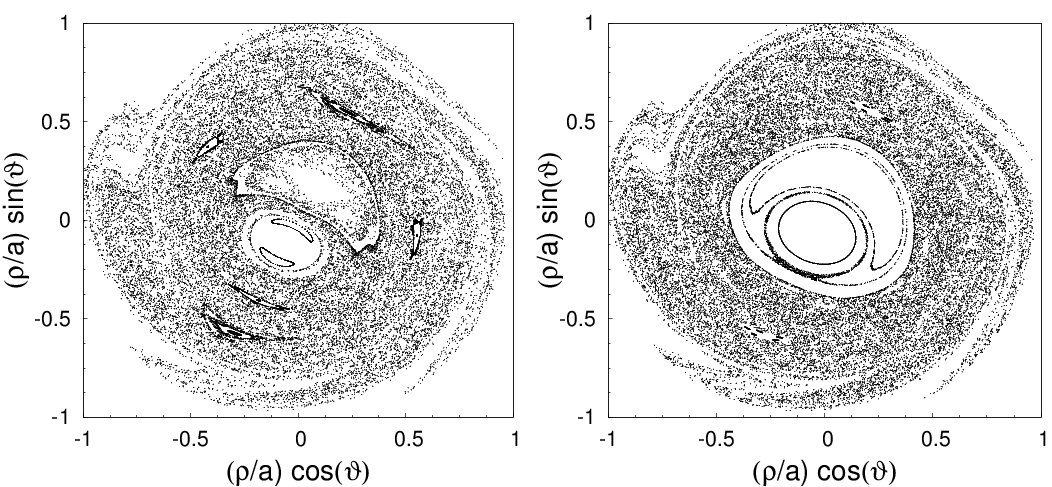} 
\caption{Poincar\'e sections of field lines in a pre--disruption plasma caused
  by several MHD modes: (a) runaway-free discharges; (b) with runaway
  electrons. The dimensionless MHD mode amplitudes are $\epsilon_{mn}=
  8.68\times 10^{-3} b_{mn}$ with  $b_{mn}=1$ for all modes in (a), and
  $b_{11}=1/4$, $b_{mn}=1$ for all other modes ($n=1,2$, $m=1-5$) in (b). The
  safety factor at the magnetic axis is $q(0)=0.8$ and at the plasma edge
  $q_a=4.7$. }       
\label{disrup_plasma_sect2}
\end{figure}
%---------------------------------------------------------------------------
In general the transport of heat and particles in the presence of RMPs 
is a three--dimensional problem. Particularly, a stochastic magnetic field
with the topological structures like ones in Figs.~\ref{disrup_plasma_sect2}
leads to poloidally and toroidally localized heat and particle deposition
patterns on wall \citep{Kruger_etal_05}. This is a general feature of 
open chaotic systems which has been observed in ergodic divertor tokamaks 
(see, e.g., \citep{Finken_etal:2005,Jakubowski_etal_06a,Abdullaev:2014}).  
The problem can be simplified when we are interested only in radial transport
rate. It can be done by introducing the radial diffusion coefficient averaged
over a poloidal angle.  

\subsection{Heat transport}
\label{heat_trans}
The electron heat conductivity in a stochastic magnetic field has been
assessed by diverse approaches. We apply here the following formula for the
electron heat diffusion $\chi_r$ deduced on the basis of simulations for 
transport of test particles, by taking into account coulomb collisions with 
background plasma species \citep{Abdullaev_13} (see also Sec.~10.4 in 
\citep{Abdullaev:2014}):  
%------------------------------------------------------------------------
\begin{equation}  \label{emp_formula}
\chi_r(\rho,T_e)= \frac{v_{\parallel} D_{FL}(\rho)}{1+L_c/\lambda_{mfp}}, 
\end{equation}
%------------------------------------------------------------------------
where $v_{\parallel} \approx v_{T_e}= 1.33\times 10^7~T_e^{1/2}$ is the thermal
velocity of electrons, $D_{FL}(\rho)$ the diffusion 
coefficient of field lines ($D_{FL}(\rho) \sim 10^{-5} \div 10^{-4}$m),
$\lambda_{mfp}=8.5\times 10^{21}~T_e^2(\rho) /n(\rho)$ the mean free path
length of electrons with the temperature $T_e$ and density $n(\rho)$ measured
in keV and $m^{-3}$, respectively, and $L_c \approx \pi q(\rho) R_0$ is the
characteristic connection length. \par   

A characteristic heat diffusion time one can estimate as $\tau_H= a^2/2\chi_r$,
where for $\chi_r$ we assume it’s magnitude at the radial position $\rho=0.566$
a. Before the disruption the local temperature here is of $0.6$~keV. This
provides $\chi_r=287$ m$^2$/s and $\tau_H=3.68 \times 10^{-4}$ s, i.e. 
of the order of the experimentally observed time for the plasma temperature
drop during the TQ after disruption. \par 

For a quantitative analysis we have modeled the time evolution of the radial
profile for the electron temperature averaged over the poloidal $\theta$ and
toroidal $\varphi$ angles, $T(\rho,t)$. This is done by solving numerically
the following diffusion equation:  
%------------------------------------------------------------------------
\begin{equation}  \label{HeatDiff_eqn}
\frac{\partial T}{\partial t} = \frac{1}{\rho} \frac{\partial }{\partial \rho}
\left[\rho \chi_r(\rho,T)\frac{\partial T}{\partial \rho} \right], 
\end{equation}
%------------------------------------------------------------------------
where the heat diffusivity is given by equation (\ref{emp_formula}) and the
applied boundary conditions are: $\partial T(\rho)/\partial \rho=0$ at 
$\rho=0$ and  $\partial T(\rho)/\partial \rho=- T/\delta_T$ at the plasma
edge $\rho=a$, where $\delta_T\simeq0.1~m$ is the characteristic 
\textit{e}-folding length for the temperature decay in the scrape-off layer.  
\par 

\begin{figure} 
\centering  (a) \hspace{5cm} (b) \\
\includegraphics[width=0.99\textwidth]{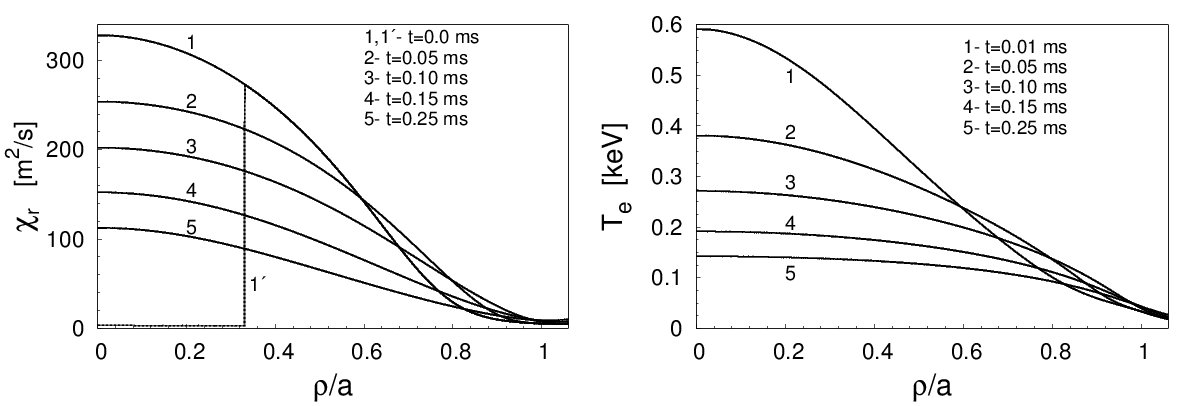} 
\caption{Radial profiles of the electron heat conductivity $\chi_r(\rho,t)$
  computed according to equation (\ref{emp_formula}), (a), and of the
  electron temperature averaged over toroidal and poloidal angles,
  $T_e(\rho,t)$, found by solving heat conduction equation
  (\ref{HeatDiff_eqn}) numerically, (b), at different time moments 
  after the disruption initiation. }    
\label{DT_t_profiles}
\end{figure}

Below we consider an example of heat transport in a fully chaotic magnetic
field shown in Fig.~\ref{disrup_plasma_sect2} (a). Figures~\ref{DT_t_profiles}
(a) and (b) show the radial profiles of the heat conductivity and the
temperature at different times. One can see in this case the
temperature drops and almost flattens within a time interval of order of 0.5
ms.   \par  
%--------------------------------------------------------------------------
%---------------------------------------------------------------------------

In the situation with partially stochastic magnetic field, see
Fig.~\ref{disrup_plasma_sect2} (b), anomalous turbulent transport in the very
plasma core, $\rho \leq 0.3$, with intact magnetic surfaces is by two orders of
magnitude smaller than in the outer region. In this case the temperature drop
in the central plasma region could be explained by the effect of parallel
transport in the magnetic island created by the $m/n=1/1$ MHD mode and some 
level of stochastization in a thin layer near the island separatrix (see a 
review by \citet{Schuller_95} for more details).  \par  

\subsection{Current quench stage}
\label{particle_trans}
As was discussed in Sec.~\ref{Experimental_evidences} the current decay rate
$\langle{dI_p/dt}\rangle$ depends on the initial RE current $I_p^{(RE)}$ which
in turn depends on the level of magnetic perturbations. Therefore, one can
assume that the current decay rate $\langle{dI_p/dt}\rangle$ may implicitly
depend on the level of magnetic perturbations. This assumption, however, may
contradict to the traditional view that the current decay rate is determined
by the time $\tau_{CQ}=R/L$, i.e., by the ratio of the plasma resistivity $R$ to
it's inductance $L$: $dI_p/dt= -I_p/\tau_{CQ}$. In the simplest case when $R$
and $L$ are constants it leads to the exponential decay law $I_p(t) \propto
\exp(-t/\tau_{CQ})$. However, in many tokamaks, particularly in the TEXTOR
tokamak, the current decay evolution is better fitted by the linear function
$I_p = a-bt$ rather than by the exponential function (see
Fig.~\ref{Ip_decay}). At present the reason of such a dependence is not
quite clear. So other mechanisms may also play role in the CQ (see, e.g.,
\cite{Kadomtsev_84,Gerhardt_etal_09,Shibata_etal_10} and references
therein). \par  

Particularly, one cannot exclude that the poloidal and toroidal variations of
the plasma current imposed by the initial MHD modes in the TQ stage do not
disappear immediately with the temperature drop (see also
Sec.~\ref{post_disr} and Eqs.~(\ref{post-disr_current}) and
(\ref{j1_prof})). This leads to the corresponding variations of the poloidal
magnetic field which acts as non--axisymmetric magnetic perturbations. Below
we discuss a possible role of the radial transport of particles in a
stochastic magnetic field created by such magnetic perturbations. \par         

The magnetic field structure before the CQ has been assumed similar to 
the one shown in Fig.~\ref{disrup_plasma_sect2}. The level of magnetic 
perturbations may be different from that during and after the thermal quench 
when strong magnetic fluctuations are present, see Fig.~\ref{119978_all} 
(d). \par  

The timescale of the current decay is determined by the rate of radial particle
transport in a stochastic magnetic field. This process has the ambipolar
nature and it is strongly collisional due to the low plasma temperature. 
On other hand one expects that the toroidal electric field induced by the
current decay also strongly affects the particle transport. Below we give a
rough estimation of the particle transport rate based on the collisional test
particle transport model. \par  
%----------------------------------------------------------------------------
\begin{table} 
  \centering
\begin{tabular} {lccc} 
$T_i$ [keV]  & $D_p$ [m$^2$/s]  & $\tau_{CQ}=a^2/2D_p$ [s] \\ [8pt]    
0.005 & 0.0986057&  1.072  \\
0.050 & 0.386249 &  2.739 $\times10^{-1}$ \\
0.100 & 1.01251  &  1.045 $\times10^{-1}$ \\
0.500 & 6.46228  &  1.637 $\times10^{-2}$ \\
1.000 & 9.51915  &  1.111 $\times10^{-2}$ \\
2.000 & 13.1030  &  8.074 $\times10^{-3}$ \\
4.000 & 17.8366  &  5.932 $\times10^{-3}$ \\
5.000 & 23.7424  &  4.456 $\times10^{-3}$ \\
10.00 & 27.0265  &  3.915 $\times10^{-3}$ \\
\end{tabular}
\caption{Ambipolar diffusion coefficients $D_p$ of particles and the
  diffusion times $\tau_{CQ}=a^2/2D_p$ from the stochastic zone at the different
  effective plasma temperatures. The plasma radius $a=0.46$ m.  } 
\label{Table_par3}
\end{table}
%------------------------------------------------------------------------- 
In Table~\ref{Table_par3} we have listed the ambipolar diffusion coefficients
$D_p$ and the characteristic diffusion times $\tau_{CQ}$ of particles at the
different plasma temperatures in a stochastic magnetic field shown in
Fig. \ref{disrup_plasma_sect2}. The typical plasma temperature after the
TQ is about from 5 eV to 50 eV. The average particle confinement
time $\tau_{CQ}$ at this temperature changes from 1 s to 0.3 s. These time
scales are shortened if the magnetic perturbation level $\epsilon^2$ is
larger. Since the diffusion coefficient $D_p \propto \epsilon^2$ and therefore
$\tau_{CQ} \propto \epsilon^{-2}$, then $\tau_{CQ}$ can be reduced to one order
smaller value for three times larger perturbation than in
Fig.~\ref{disrup_plasma_sect2}. This timescale is still much longer than the
experimentally observed values. However, this collisional model does not takes
account the effect of the inductive toroidal electric field. One expects that
the acceleration of electrons and ions by the electric field increases
the radial transport of particles. To include this effect in the collisional
model one can assume that the effective temperature of the plasma is higher
than the measured one. The particle diffusion time $\tau_{CQ}$ for the effective
temperature 2 keV is about $8\times 10^{-3}$ s. This timescale gives the
average current decay rate $|dI_p/dt| \approx I_p/\tau_{CQ}= 0.35/(8.0 \times
10^{-3}) \approx 44.0$ MA/s which is of the order of the experimental measured 
rates given in Table~\ref{Table_Ipt}.   \par    

More rigorous approach to the particle transport in a stochastic magnetic
field during the current decay stage would require a three--dimensional
treatment of the problem. It should take into account not only the formation of
the ambipolar electric potential \citep{Spizzo_etal_14} but also the inductive
toroidal electric field which accompanies the process. The latter may lead to
the directionality of the particle transport that eventually may influence on
the random vertical displacement of runaway beams. Of course, the study of
these complicated processes is out of the scope of the present work.  \par

%---------------------------------------------------------------------------
\section{Modeling of post--disruption plasma}
\label{post_disr}
The described scenario of plasma disruption with a RE beam allows one to
model a post-disruption plasma. After establishing the runaway beam the
current is localized inside the area enclosed by the last intact magnetic
surface. In general the distribution of the current density $j$ would depend
not only on the radial coordinate $\rho$ but also vary along the poloidal
$\theta$ and the toroidal $\varphi$ angles due to the presence of the
($m/n=1/1$) magnetic island. Such a post--disruption plasma current can be
presented as a sum of two parts,  
%----------------------------------------------------------------------------
\begin{equation} \label{post-disr_current}
j(\rho,\theta,\varphi)= j_0(\rho) +j_1(\rho,\theta,\varphi),
\end{equation}
%----------------------------------------------------------------------------
where $j_0(\rho)$ is the current density depending only on the radial
coordinate $\rho$, and $j_1(\rho,\theta,\varphi)$ is the helical current which
is a periodic function of the poloidal $\theta$ and toroidal $\varphi$
angles. \par  

The radial dependence of $j_0(\rho)$ can be also modeled by assuming that
after the disruption the current is uniformly distributed over confined area
with the steep gradient at the beam edge $\rho_c$. Calculations show that
electron orbits does not significantly depend on the specific of the radial
profile of $j_0(\rho)$. For our calculations of GC orbits we choose the
following profile 
%----------------------------------------------------------------------------
\begin{equation} 
j_0(\rho) = 
\left\{ \begin{array}{ll}
\displaystyle J_0 \tanh \left[(\rho_c^2-\rho^2)/\Delta_a\right], \quad
\mbox{for\ } \quad \rho <\rho_c, \\ [8pt] 
\displaystyle 0,  \quad \mbox{ for\ } \quad  \rho > \rho_c,
\end{array} \right.
\label{pd_current_prof}
\end{equation}
%----------------------------------------------------------------------------
where $J_0$ is the constant determined by the full current of the beam
$I_p^{(RE)}$ and $\Delta_a$ is the steepness parameter. The current flowing
inside magnetic surface $\rho$, i.e., $I_p(\rho) = 2\pi J_0
\int\limits_0^{\rho} j_0(\rho') \rho' d\rho'$ is given by 
%---------------------------------Eq.()---------------------------------
\begin{equation}   \label{Ip0_tanh}
I_p(\rho) = 
\left\{ \begin{array}{ll}
\displaystyle I_{p}^{(RE)} \left[1 - \frac{\ln
    \cosh\left([\rho_c^2-\rho^2]/\Delta_a\right)}{\ln
    \cosh\left(\rho_c^2/\Delta_a\right)} \right], \quad \mbox{for\ } \quad \rho
\leq \rho_c, \\ [8pt]    
\displaystyle I_p^{(RE)}, \quad \mbox{for\ } \quad \rho > \rho_c,
\end{array} \right.
\end{equation}
%----------------------------------------------------------------------------
where $I_{p}^{(RE)}$ is the full current of the confined area.  \par

One should also note the fact that after the TQ the plasma beam is
shifted inwardly because of drop of plasma pressure. In the modeling this
fact can be taken into account by assuming that the radial position of the
center $R_a$ of the post--disruption plasma is different from the one of the
pre--disruption plasma. The safety factor of the corresponding plasma is then
given by 
%---------------------------------Eq.()---------------------------------
\begin{eqnarray}   \label{q_p0_tanh}
& q(\rho)= q_{cyl}(\rho) C(\rho/R_a), \hspace{1cm} 
q_{cyl}(\rho) = \dfrac{2\pi\rho^2B_0}{\mu_oR_aI_p(\rho)},  
\end{eqnarray}
%----------------------------------------------------------------------------
where $q_{cyl}(\rho)$ is the safety factor of the cylindrical plasma, the
function $C(x)=1+A_1 x + A_2x^2 + \cdots$ is a function which takes into
account the toroidicity of plasma. The coefficients $A_i$, ($i=1, 2, \dots$)
depends on the plasma pressure \citep{Abdullaev_etal_99,Abdullaev:2006}. \par 
 
\begin{figure} 
\centering  
\includegraphics[width=1.0\textwidth]{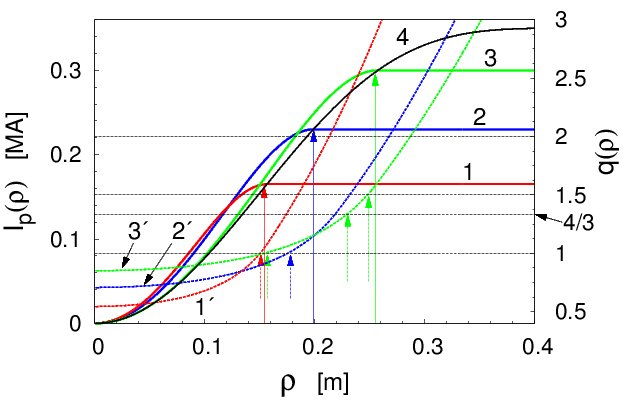}
\caption{Radial profiles of the plasma current $I_p(\rho)$ (\ref{Ip0_tanh})
  (solid curves 1$-$3 on l.h.s. axis), the safety factor profiles $q(\rho)$
  (\ref{q_p0_tanh}) (dashed curves $1'-3'$ on r.h.s. axis), and curve 4
  corresponds to the pre--disruption plasma current. The red curves 1 and 1'
  correspond to $I_p^{(RE)}=165$ kA, blue curves 2 and 2' correspond to
  $I_p^{(RE)}=230$ kA, and green curves 3 and 3' correspond to $I_p^{(RE)}=300$
  kA. The vertical solid arrows indicate the radii of plasma beam $a$, the
  vertical dashed arrows indicate the positions of resonant magnetic surfaces
  $q=1$ and $q=3/2$. The toroidal magnetic field $B_t=2.4$ T, $R_a=1.7$ m, the
  pre--disruption plasma current $I_{p}=350$ kA and the radius $a_0=0.46$ m. }
\label{post_plasma}
\end{figure}

Figure~\ref{post_plasma} shows the radial profiles of $I_p(\rho)$ (solid
curves 1--3 on the l.h.s. axis) (\ref{Ip0_tanh}) and the safety factor
(\ref{q_p0_tanh}) (dashed curves $1^\prime - 3^\prime$ on the r.h.s. axis) for the
three discharge parameter, respectively. Solid black curve 4 corresponds to the
pre--disruption plasma current profile. We set the toroidal field magnitude
$B_0=2.4$ T, the beam center at $R_a=1.7$ m. The plasma radius $a$ is found
from the condition $I_{p0}(a)=I_p^{(RE)}$ where $I_{p0}(r)$ is the current
profile of the pre--disruption plasma. The vertical dashed color arrows show
the radial positions of the $q=1$, $q=4/3$, and $q=3/2$ magnetic surfaces and
the vertical solid arrows indicate the plasma radii $a$.  \par 
%--------------------------------Fig.-------------------------------------
%-------------------------------------------------------------------------
Note that that the red curves 1 and $1^\prime$ and green curves 3 and $3^\prime$ 
in Fig.~\ref{post_plasma} correspond to the discharges with the lowest and
highest values of $I_p^{(RE)}$ shown in Fig.~\ref{119978_all}~(b). For the 
lowest value of $I_p^{(RE)}$ the radial position of the $q=1$ magnetic surface 
is very close to the RE beam radius $\rho_c$. For the highest value of 
$I_p^{(RE)}$ the magnetic surfaces with $q=1$, $q=4/3$, and $q=3/2$ are located 
inside the plasma region $\rho < a$. However, the radial position of the 
magnetic surface $q=3/2$ is at the plasma edge. For the typical discharges
like the one shown by blue curves the magnetic surface $q=1$ is located
relatively far from the plasma edge. \par    

The Fourier expansion of the helical current, $j_1(\rho,\theta,\varphi)$, 
%----------------------------------------------------------------------------
\begin{equation} \label{j1_prof}
j_1(\rho,\theta,\varphi) = \sum_{m,n} j_{mn}(\rho) \cos(m\theta -n\varphi+
\phi_{mn}), 
\end{equation}
%----------------------------------------------------------------------------
is mainly dominated by the $m/n=1/1$ component. This assumption is based on
the analysis of numerous disruptions in the JET tokamak
\citep{Gerasimov_etal_14}.   \par 

We should assume that the value of the safety factor at the beam axis $q(0)$
is less than unity. This assumption is supported by a number of experimental
measurements of the current profile after the sawtooth crashes in the TEXTOR,
the TFTR, and JET tokamaks
\citep{Soltwisch_etal_87a,Yamada_etal_94,SoltwischKoslowski_95,% 
ORourke_91,Koslowski_etal_96,SoltwischKoslowski_97}.    \par   

This model of the post--disruption plasma current describes only the initial
stage of RE beam. During acceleration of electrons in the toroidal electric
the RE orbits drift outward and their form evolves from the circular one to
oval one. This process changes in turn the RE beam form and its current. The
self--consistent description of the time evolution of the RE beam is a
difficult problem. It is beyond of the scope of this present study. \par     

%----------------------------------------------------------------------------
\section{Evolution of GC orbits during acceleration} 
\label{evolution_REs}
Now we discuss the dynamics of RE orbits during the acceleration of electrons
induced by the toroidal electric field in a toroidal post--disruption plasma.
Mainly we study the outward drift of RE orbits and it's role in the RE
losses. However, we will not consider the processes of the generation and
the proliferation of the RE population, the problems of stability of RE beams
and related issues. These problems have been much discussed in literature.  

\subsection{Outward drift of RE orbits}
\label{Outward_drift_RE_orbits}
\begin{figure} 
\centering  (a) \hspace{5cm} (b)  \\
\includegraphics[width=0.99\textwidth]{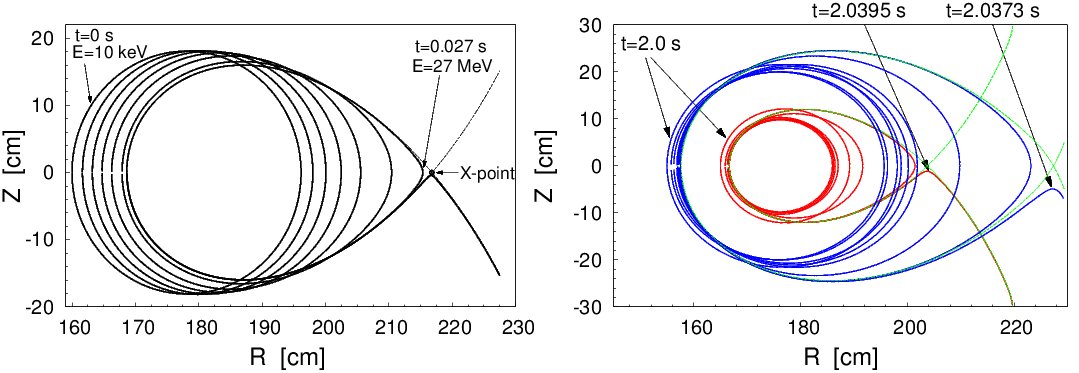}
\caption{(a) Evolution of the GC orbit of accelerating electron in the
  ($R,Z$)-plane at the constant plasma current $I_p=100$~kA in the presence
  of a constant toroidal electric field with  $V=40$ V. Dashed curve 
  corresponds to the separatrix of GC orbits of electron of energy $E=27$ MeV.
(b) The same as in (a) but for the two GC orbits
  of accelerating electrons for the time-- varying current $I_p(t)$ and the
  loop voltage $V(t)$ corresponding to the TEXTOR discharge \#117527. 
  Green curves correspond to the separatrices of GC orbits. }
\label{Ip_orbits_const}
\end{figure}

First we consider the case of the axisymmetric plasma beam neglecting the
helical magnetic perturbations. The inductive toroidal electric field
generated due to the current decay during the plasma disruption accelerates
thermal electrons. This is an adiabatic process since the characteristic time
of significant variation of energy is much larger than the transit time of
electrons. Therefore the GC orbit slowly drifts outward without changing the
area of GC orbit in the poloidal plane which is an adiabatic invariant $J$
or the action variable (see Sec.~6.1 of Supplementary part)). With increasing
electron energy the topology of GC orbits also slowly changes from the
circular one to the oval one. Starting from the certain critical energy
$E_{cr}$ the adiabaticity of the process breaks and the GC orbit bifurcates by
creating the unstable stagnation point (or X-points) inside the plasma
region. With the further increase of energy the GC orbit crosses the
separatrix (a homoclinic orbit associated the X-point) and becomes
unconfined. The value $E_{cr}$ depends on the plasma current $I_p$. The
described phenomenon is an addition mechanism of confinement loss of
REs. Figure~\ref{Ip_orbits_const} (a) shows a typical evolution of a GC orbit
in the presence of the toroidal electric field with the constant beam
current $I_p=100$ kA and the loop voltage $V=40$ V. \par

One should note that the formation of the separatrix of RE GC orbits during 
the acceleration process in tokamaks has been first predicted in
Ref.~\citep{Zehrfeld_etal_81}. The numerical study of this process in a
realistic tokamak configuration has been carried out in
Ref.~\citep{Wongrach_etal_14}. Particularly, it was shown that with increasing
electron energy the area confined by the separatrix decreases and it vanishes
when the energy exceeds a certain critical value $\mathcal{E}_{cr}$, i.e. such
electrons cannot be confined. The critical energy $\mathcal{E}_{cr}$ is
proportional to the square root of the plasma current $I_p$, $\mathcal{E}_{cr}
\propto \sqrt{I_p}$. \par  

The described evolution of RE orbits is in agreement with the experimental
observation of the IR radiation patterns observed in the experiment in the
TEXTOR tokamak \citep{Wongrach_etal_14} and in the DIII-D tokamak
\citep{Hollmann_etal_13}. The observations clearly show the evolution of the
spatial form of RE beam from crescent ones into oval ones with increasing the
electron energies. \par   

The example of the time--evolution of GC orbits in the plasma beam with a
time--varying current $I_p(t)$ and the loop voltage $V(t)$ corresponding to
the TEXTOR discharge \#117527 is shown in Fig.~\ref{Ip_orbits_const} (b). To
simplify the calculations of orbits we have assumed that the loop voltage
$V(t)$ is uniform in the poloidal section, i.e., it does not depend on the
radial coordinate $r$ and equal to the experimentally measured value at the
limiter. However, this assumption only approximately describe the
situation. To find more the exact magnitudes of the toroidal electric field
during the runaway current decay one should solve the corresponding Maxwell
equations.     \par    
%--------------------------------Fig.2-------------------------------------
\begin{figure} 
\centering 
\includegraphics[width=0.49\textwidth]{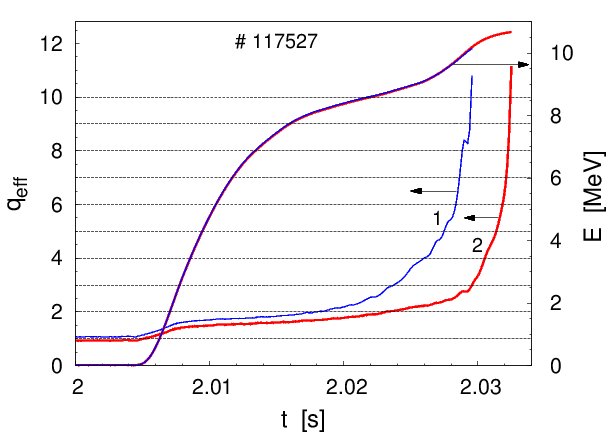}
\caption{Time--evolution of the effective safety factors $q_{eff}$
  (l.h.s. axis) and electron energies (r.h.s. axis) during the acceleration in
  the discharge  \#117527. Blue curves correspond to the orbit launched at the
  coordinate $(R=160,Z=0)$ cm, and red curves to the one with  $(R=165,Z=0)$
  cm. Horizontal lines correspond to $q(t)=m$ where $m=1, 2, \dots$ are the
  integer numbers.}    
\label{q_eff_time}
\end{figure}

%-------------------------------------------------------------------------
One of the important parameter of the GC orbit is the effective safety factor
$q_{eff}$ defined as a ratio $q_{eff}=\Delta\varphi /2\pi$ where
$\Delta\varphi$ is the increment of the toroidal angle $\Delta$ per one 
poloidal turn. It is a function of the action variable $J$ and particle energy
$E$. For low--energy electrons the quantity $q_{eff}(J,E)$ coincides with the
safety factor $q(\rho)$ of the equilibrium magnetic field. With increasing the
electron energy the effective safety factor strongly deviates from
$q(\rho)$. With approaching RE energy $E$ to the critical one $E_{cr}$ it
diverges as   
%-------------------------------Eq.()--------------------------------------
\begin{equation} \label{asym_q_eff}
q_{eff}(J,E) \propto -\ln |E-E_{cr}|.
\end{equation}
%-----------------------------------------------------------------------
Figure~\ref{q_eff_time} shows the typical time evolutions of the effective
safety factors $q_{eff}$ of two GC orbits during the electron acceleration in
the conditions of the TEXTOR discharge \#117527. \par  
%--------------------------------------------------------------------
%--------------------------------------------------------------------

\begin{figure} 
\centering 
\includegraphics[width=0.6\textwidth]{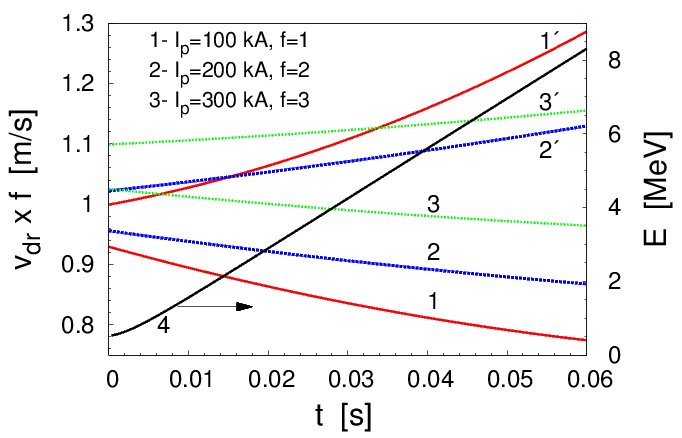} 
\caption{Drift velocities of innermost $v_{df}(R_{i})$ (curves 1, 2, and 3) and
  outermost $v_{df}(R_{o})$ (curves 1$^\prime$, 2$^\prime$, and 3$^\prime$) points
  of orbits for the different plasma current: curves 1 and 1$^\prime$
  correspond to the plasma current $I_p=100$ kA, curves 2 and 2$^\prime$
  correspond to $I_p=200$ kA, and curves 3 and 3$^\prime$ correspond to
  $I_p=300$ kA. Curve 4 describes the increase of energy $E$ (right hand
  axis). The toroidal field $B_t=2.5$ T,  major radius $R_0=175$ cm, minor
  radius $a=$46 cm, the loop voltage $V_{loop}=5$ V. Note that $v_{dr}$ is
  multiplied to the proportionality factor $f$. }     
\label{orb_drifts}
\end{figure}

Figure \ref{orb_drifts} shows the time--evolution of the outward drift
velocity $v_{dr}$ calculated numerically for the three different RE beam
currents. It is quite well described by formula derived in
Ref.~\citep{Abdullaev_15}) (see also Sec.~4 of Supplementary part)   
%-----------------------------------------------------------------
\begin{align} \label{Delta_v_RT}
v_{dr}=  \dfrac{R_0E_\varphi}{RB_z^*} \left(1-\frac{RT_{av}}{R_0T} \right),     
\end{align}
%------------------------------------------------------------------
where $B_z^* =B_z + F(E)$ is the effective poloidal magnetic field, $B_z$ is
the $z-$ component of the poloidal magnetic field at the equatorial plane
$z=0$, $F(E)$ is the term depending on a particle energy. The quantity 
%------------------------------------------------------------------
\begin{align} \label{B-pol} 
& T_{av}= \frac{2\pi q_{eff}R_0}{v_\varphi},
\end{align}
%------------------------------------------------------------------
is the average transit time, $E_\varphi$ is the toroidal electric field
strength, $T$ is the transit time of orbit, $v_\varphi$ is the toroidal
velocity. \par   

The expression (\ref{Delta_v_RT}) describes the creation of the X-point and
the separatrix of RE orbits at the critical energy $E_{cr}$. This phenomenon is
related with the appearance of zeroes of the effective poloidal magnetic field 
$B_z^*$ at  $E=E_{cr}$ and the certain radial distance $R=R_s$ within the
plasma region (see \citet{Abdullaev_15} for details). \par  

At $|B_z| \gg |F(E)|$, and $T_{av} \approx T$ the formula (\ref{Delta_v_RT})
is reduced to 
%-----------------------------------------------------------------
\begin{equation} \label{drift_v_Guan}
v_{dr}=  \frac{qE_\varphi}{B_0}= - \frac{(R-R_0)E_\varphi}{RB_Z} ,      
\end{equation}
%------------------------------------------------------------------
obtained by \citet{Guan_etal_10,Qin_etal_11} for the circular orbits. Here
$q=(R-R_0)B_0/B_zR$ is the safety factor of magnetic field. As seen from
Fig.~\ref{orb_drifts} the formulas (\ref{Delta_v_RT}) and (\ref{drift_v_Guan})
give the correct dependence of $v_{dr}$ on the plasma current $I_p$, $v_{dr}
\propto I_p^{-1}$ because $B_z \propto I_p$.      \par     
%----------------------------------------------------------------------------
%----------------------------------------------------------------------------
However, the formula (\ref{drift_v_Guan}) does not describe the situation when
the GC orbits take an oval form with increasing the energy similar to the ones
shown in Figs.~\ref{Ip_orbits_const} (a) and (b). From the latter it
follows that the average outward velocity $v_{dr}$ of the innermost part of the
orbit is approximately equal 0.6 m/s and 8 m/s of the outermost part of the
orbit.  \par   

\subsection{RE current decay}
The rate $dI_p/dt$ of the runaway current loss due to described outward drift
of orbits can be roughly estimated as follow. This loss mechanism is mainly
caused by the shrinkage of the beam radius $\rho_c$. The rate of such a
shrinkage $d\rho_c/dt$ is of order of the average outward velocity
$v_{dr}$. Since $I_p\propto \rho_c^2$, we have 
%----------------------------------------------------------------------------
\begin{equation}  \label{current_dr}
\frac{dI_p}{dt} = \frac{dI_p}{d\rho_c}\frac{d\rho_c}{dt} \sim \frac{2I_p}{\rho_c}
v_{dr} \propto \frac{E_\varphi}{\rho_c}.  
\end{equation}
%----------------------------------------------------------------------------
For the typical values of $I_p\approx 0.2 MA$, $\rho_c \approx 0.2$ m, and
$v_{dr}\sim 1$ m/s one has $dI_p/dt \approx 4$ MA/s. This estimation is of
order of the experimentally measured average decay rate of the runaway current
listed in Table \ref{Table_Ipt}. \par     

Since the safety factor $q$, as well as $q_{eff}$ of RE beams is about unity,
$q_{eff} \sim 1$ then the outward drift may slowdown for the higher values of
the toroidal magnetic field $B_0$. However, much the higher toroidal electric
field $E_\varphi$ may compensate this effect so that the decay time of RE
currents in large tokamaks, like ITER, may have the same order as in smaller
tokamaks.   \par    

One should also note that the outward drift velocity $v_{dr}$ is proportional
to the inverse aspect ratio of tokamaks, $v_{dr} \propto a/R_0$
\citep{Abdullaev_15}. It means that the RE current loss due to the outward
drift of orbits in spherical tokamaks would be larger than in standard
tokamaks, so that the RE electron would cease faster. This effect could be one
of reasons of the absence of REs during disruptions in NSTX tokamaks
\citep{Gerhardt_etal_09}.      

Beside of outward orbit drifts the RE current losses is also caused by the
internal MHD mode which be discussed in the next section. The collisions of
REs with neutral particles may also contribute to the RE losses. 
%--------------------------------------------------------------------

\section{Effect of magnetic perturbations} 
\label{Helical}
The effect of the magnetic perturbations on electrons in the 
post--disruption current beam strongly depends on its safety factor profile
$q(\rho)$, the spectrum of magnetic perturbations, and the electron
energy. To explain this effect we consider the simplified version of GC motion
equations in the presence of magnetic perturbations. (The rigorous
consideration of this problem is given in Sec.~6 of Supplementary Part). \par

The particle drift motion in the presence perturbations can be presented by
Hamiltonian equations similar to the equations for magnetic field lines, 
%-------------------------------Eq.()--------------------------------------
\begin{align} \label{Heqn_action}
& \frac{d\vartheta}{d\varphi} =  \frac{\partial K}{\partial J}, & 
\frac{dJ}{d\varphi} =-\frac{\partial K}{\partial \vartheta_z},  
\end{align}
%-----------------------------------------------------------------------
with the Hamiltonian $K= K(\vartheta,J,E, \varphi)$ with the canonical
variables $(\vartheta, J)$, and the toroidal $\varphi$ as the time--like
variable. In the absence of perturbations GC orbits wound the drift surfaces
$J=$const and the poloidal angle $\vartheta$ is a linear function $\varphi$,
$\vartheta=\varphi/q_{eff}(J,E)+\vartheta$. In the presence of perturbations
Hamiltonian $H$ can be presented as a sum  
%-------------------------------Eq.()--------------------------------------
\begin{align} \label{H01_action}
K= \int \frac{dJ}{q_{eff}(J,E)} + \epsilon K_1(\vartheta_z, J,E, \varphi).   
\end{align}
%-----------------------------------------------------------------------
Since the perturbation are periodic in poloidal and toroidal angles and in
time it can be presented by a Fourier series 
%-------------------------------Eq.()--------------------------------------
\begin{align} \label{H1_series}
K_1(\vartheta_z, J, E, \varphi) =  \sum_{mn}
K_{mn}(J,E) \exp\left[i \left(m\vartheta-n \varphi \right) \right].   
\end{align}
%-----------------------------------------------------------------------
The strongest influence of perturbation on particles takes place on the
($m,n$) resonant drift surfaces, i.e.,   
%-------------------------------Eq.()--------------------------------------
\begin{align} \label{res_condition_q}
m = nq_{eff}(J,E),
\end{align}
%-----------------------------------------------------------------------
originating from the ($m,n$) term in (\ref{H1_series}) with the amplitude
$K_{mn}(J,E)$. They are determined by the magnetic perturbation spectrum
$b_{mn}$, 
%-------------------------------Eq.()--------------------------------------
\begin{align} \label{Hmn_b_mn}
K_{mn}(J,E) \propto  \sum_{m'} b_{mn} \int_0^{2\pi} d\vartheta 
\exp\left[i \left(m\vartheta-m' \vartheta_M \right) \right],   
\end{align}
%-----------------------------------------------------------------------
where $\vartheta_M$ the poloidal angle associated with magnetic field lines is
a function of $\vartheta$ as well as particle energy $E$. \par  

\begin{figure} 
\centering   (a) \hspace{5cm} (b) \\
\includegraphics[width=0.99\textwidth]{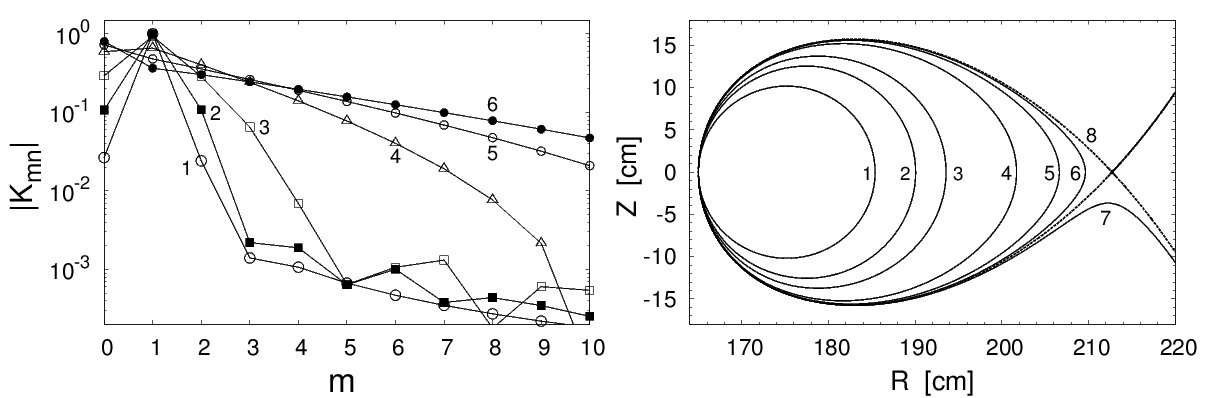}
\caption{(a) Spectrum of perturbations $K_{mn}$ and (b) corresponding RE
  orbits with different energies $E$. Curves 1--7 correspond 
to RE energies 10 keV, 20 MeV, 30 MeV, 40 MeV, 42 MeV, 42.5 MeV, and
42.7 MeV, respectively. Curve 8 corresponds to the separatrix with the
critical energy $E_{cr}=42.646$ MeV. The plasma current $I_p=$ 150 kA, the
toroidal field $B_0=2.5$ T. The toroidal mode number $n=1$. }   
\label{K_mn_vs_energy}
\end{figure}

For {\em low--energy electrons} (up to 5 MeV) the spectrum of amplitudes
$K_{mn}(J,E)$ weakly depends on energy $E$ and close to the spectrum
of magnetic perturbations $b_{mn}$ of $(m,n)-$th modes. With increasing the
energy the spectrum of perturbations $K_{mn}(J,E)$ deviates from
$b_{mn}$ and acquires more higher poloidal harmonics $m$. The example of the
poloidal spectra of perturbation $K_{mn}(J,E)$ for different particle
energies is shown in Fig.~\ref{K_mn_vs_energy} (a). The corresponding
unperturbed orbits are plotted in Fig.~\ref{K_mn_vs_energy} (b).  It is
assumed that the magnetic perturbation contains a single $(m=1,n=1)$ mode. 
%----------------------------------------------------------------------------
%----------------------------------------------------------------------------
For the low energy electrons with $E<10$ MeV the spectrum $K_{mn}$ contains
the predominant $m=1$ mode. \par 

With increasing the energy the amplitudes $K_{mn}$
of higher $m$ also grow and the width of the poloidal spectrum $K_{mn}$ in 
$m$ becomes wider as shown in Fig.~\ref{K_mn_vs_energy} (b). For the spectrum
$K_{mn}$ one can obtain the following asymptotical formula for the orbits
close to the separatrix (see Sec.~3.4 in  \citep{Abdullaev:2014}) 
%-------------------------------Eq.()--------------------------------------
\begin{align} \label{asym_Hmn}
&K_{mn} \propto \frac{1}{q_{eff}} \exp\left(-\frac{mC}{q_{eff}} \right),  
\end{align}
%-----------------------------------------------------------------------
where $C$ is a finite constant, and the effective safety factor $q_{eff}$
diverges as (\ref{asym_q_eff}). \par 

As was shown in Sec.~\ref{post_disr} (see also Fig.~\ref{post_plasma}) the
typical values of $q(\rho)$ varies between $q(0)\approx 0.7 \div 0.8$ at the
magnetic axis and $q(a) < 1.5$ at the plasma edge. Therefore, the strongest
effect of the RMPs on electron orbits may expect if its spectrum $b_{mn}$
contains a sufficient number of $(m,n)-$ components that are resonant to the
magnetic surfaces with $q$ in the interval $q(0) <q=m/n < q(a)$ that would
create a stochastic zone of magnetic field lines. The electrons from this
stochastic layer would be then radially transported to wall. \par   

Below we discuss the influence of magnetic perturbations on RE orbits for the
two specific cases. First we consider the effect of internal single helical
magnetic field, and then we analyze the effect of the external RMPs, namely,
the TEXTOR-DED on the confinement of REs.   

\subsection{Effect of a single helical magnetic field}
\label{helical_pert}
\begin{figure} 
\centering  (a) \hspace{5cm} (b) \\
\includegraphics[width=0.99\textwidth]{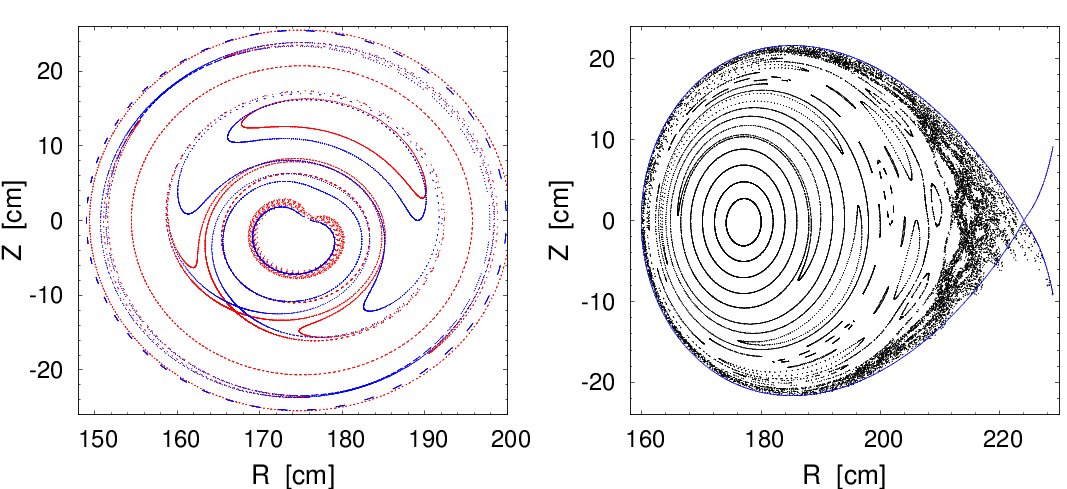}
\caption{Poincar\'e sections of RE orbits in the ($R,Z$)-plane: Red dots
  correspond to RE orbits with energies $E=1$ MeV (a) and $E=20$ MeV (b) ,
  blue dots correspond magnetic field lines (a); green curve is the
  separatrix. The perturbation parameter $\epsilon=10^{-5}$, the plasma
  currents $I_p=200$ kA (a) and  $I_p=100$ kA (b). } 
\label{fig_orbits_Energy}
\end{figure}

Assume the magnetic perturbation (\ref{A_MHD_sum}) contains the single
$(m=1,n=1)$ MHD mode as was proposed in the model of the post--disruption
current beam described in Sec.~\ref{post_disr}. For the low--energy electrons
it creates a single island structure since the deviations of their GC orbits
from the magnetic surfaces is small. Such a {\em system is stable} because the
single MHD mode does not create stochasticity of magnetic field lines. The
example of this case is shown in Fig.~\ref{fig_orbits_Energy} (a) by the
Poincar\'e sections of RE orbits (red dots) and magnetic field lines (blue
dots). \par   

With increasing the energy of electrons and decreasing the beam current the
electron's GC orbits strongly deviate from the magnetic field lines. The
effective safety factor $q_{eff}$ of the GC orbit increases as the electron
energy grows as was shown in Fig.~\ref{q_eff_time}. At certain time instants
the value of $q_{eff}$ reaches the integer value so that the resonant
condition may be satisfies for the
higher harmonics $(m>1,n>1)$ of the GC orbits with the $(m=1,n=1)$ magnetic
perturbation. This generates a number of island chains of GC orbits. The 
interaction of several such island structures may even lead to the formation
of the stochastic layer near the separatrices [see 
Figs.~\ref{Ip_orbits_const} (a) and (b)].  
%--------------------------------------------------------------------
%--------------------------------------------------------------------

Figure~\ref{fig_orbits_Energy} (b) illustrates the typical structure of
high--energy electrons in the presence of the internal helical magnetic field
with a single $(m=1,n=1)$ mode. Such a structure  
leads to the widening area of lost electrons and decreasing the critical
energy $E_{cr}$.  The characteristic escape time of REs from the stochastic
layer is of order of 10 $\mu$s. Sudden RE bursts in many discharges is
probably related with the loss REs from the stochastic layer. Occurrence of
the MHD mode signals accompanied these events will be discussed in the next
Sec.~\ref{MHD_mode_gen}.      \par 

As was discussed above in Secs.~\ref{description} and \ref{post_disr} (also
Figs.~\ref{119978_all}~(b) and \ref{post_plasma}) there are some exceptional
discharges (for example \#117859) with the highest RE current and several
low--order rational surfaces within plasma beam. Such a beam can be
easily destabilized by the magnetic perturbations containing several MHD
modes with low--order ($m,n$) numbers. Such a magnetic perturbation may affect
strongly on electrons creating the chaotic zone at the beam edge
open to wall. Such an effect probably explains the sudden lost of REs at
certain times seen in Fig.~\ref{119978_all}~(b).         

\subsection{Influence of the TEXTOR--DED}
\label{Effect_DED}
The coil configuration of the TEXTOR--DED is designed to have the poloidal
spectra of magnetic perturbations localized near the magnetic surface $q=3$ of
the flat--top plasma discharges (see Sec.~5.2 of Supplementary
part). Therefore, these perturbations do not contain a necessary number of
resonant components to create a stochastic zone of magnetic field lines in the
post--disruption current beam with the safety factor $q$ lying between $q(0)
<1$ and $q(a)<1.5$.  \par 

In the so--called 3/1 operational mode with the predominant toroidal mode
$n=1$ toroidal there is only one $(m=1,n=1)$ component resonant to the
magnetic surface $q=1$. The similar situation takes place in the 6/2 mode
($n=2$) with the resonant component $(m=2,n=2)$. [There are no magnetic
surfaces in the plasma region that are resonant to the components $(m=1,n=2)$
and $(m=3,n=2)$]. On the other hand this resonant component of the DED field
is weak since it is located away from the maximum of the spectrum.  Therefore,
the effect of the DED on the RE beam does not create the stochastic zone of
{\em magnetic field lines} from which electrons would escape to wall as in the
case of the stochastic zone in a flat--top plasma operation. The only
$m/n=1/1$ component of the DED perturbations may create an 
island structure near the $q=1$ magnetic surface similar to one shown in
Fig.~\ref{fig_orbits_Energy} (a). \par 

\begin{figure}   
\centering (a) \hspace{5cm} (b) \\
\includegraphics[width=0.99\textwidth]{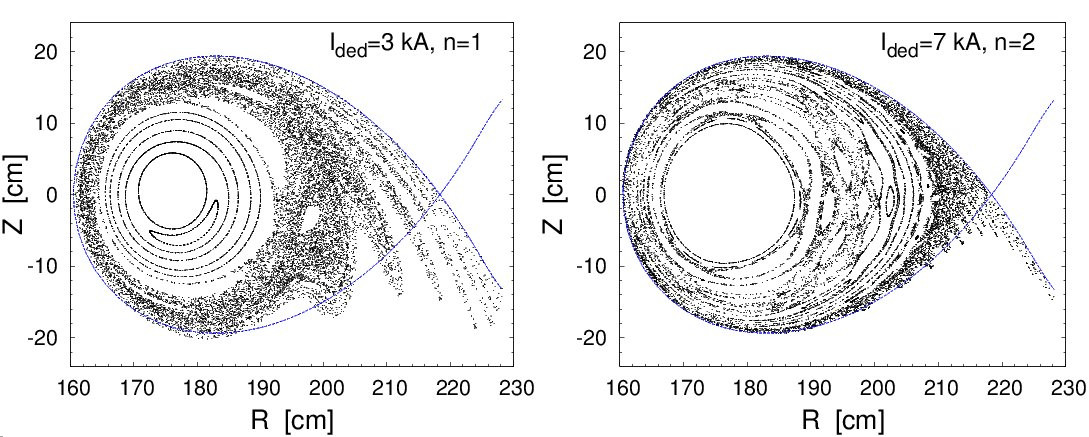}
\caption{Poincar\'e sections of RE orbits in the ($R,Z$)-plane of the RE
  orbits of energy $E=20$ MeV in the presence of the magnetic perturbations of
  the TEXTOR-DED. (a) corresponds to the $n=1$ mode and the DED current
  $I_{ded}=3$ kA; (b) corresponds to the $n=2$ mode and $I_{ded}=7$ kA. The
  plasma current $I_p=94$ kA, the toroidal field $B_0=2.4$ T. }    
\label{fig_orbits_DED}
\end{figure}

With increasing the energy of REs and decreasing the plasma current the
DED perturbation starts to affect on REs because of appearance of high--mode
resonances $q_{eff}=m/n$ similar to the case discussed in
Sec.~\ref{helical_pert}. It generates the structures with islands and a
stochastic layer. Figures~\ref{fig_orbits_DED} (a) and (b) show the typical
Poincar\'e sections of GC orbits of energetic electrons affected by the
TEXTOR--DED: (a) corresponds to the $3/1$ mode with the DED current
$I_{ded}=3$ kA; (b) corresponds to the $6/2$ mode with $I_{ded}=7$ kA. The
particle energy is taken $E=20$ MeV, the plasma current $I_p=94$ kA, and the
toroidal field $B_0=2.5$ T. These structures explain the fast decay of RE
current in its final stage accompanied by spikes in the scintillation probe
(see Fig.~\ref{119978_all} (b)). \par  
%------------------------------------------------------------------------
%------------------------------------------------------------------------
The structures of RE orbits shown in Figs.~\ref{fig_orbits_Energy} (b) and
\ref{fig_orbits_DED} correspond to 
the final termination stages of RE current. They have features which are 
characteristic for the so--called stable and unstable manifolds created by the
splitting of separatrices (see, e.g., \citep{Abdullaev:2014}). They lead to the
toroidally and poloidally localized deposition patterns of REs on wall. 
Toroidal peaking and spatial--temporal evolution of hard X-ray emission in the
final stage of RE current loss observed in DIII-D experiments
\citep{James_etal_12} is consistent the described topology of REs.  \par

The experimental observations in the TEXTOR-DED have indeed showed that the
RMPs field which switched on just after the TQ does not affect on
the radial transport and the loss of low--energy electrons
\citep{Koslowski_etal_14,Wongrach_etal_15} (see also
Table~\ref{Table_Ipt}). This is mainly because of the mentioned features of
the poloidal and toroidal spectra of the DED field. 

\subsection{Generation of magnetic perturbations by high--energy electrons} 
\label{MHD_mode_gen}
The above mentioned in Sec.~\ref{description} the occurrence of the MHD
activities during the sudden RE bursts can be explained by the nonlinear
interaction of high--energy electrons with the $(m=1,n=1)$ MHD mode. The MHD
magnetic perturbations with mode numbers $(m,n)-$ higher than the initial
$(m=1,n=1)$ mode can be generated during the acceleration process of the
REs. At certain energy of REs their orbits strongly deviate from the magnetic
surfaces which creates in turn higher $(m,n)-$harmonics, $(m>1,n>1)$, of the
MHD ($m=1,n=1$) mode (\ref{Hmn_b_mn}). The resonant interaction of RE orbits
with these harmonics leads to the redistribution of corresponding current near 
these orbits according to the helicity of these modes. Therefore, the current
density (\ref{j1_prof}) acquires higher $(m,n)-$components $j_{mn}$ which in
turn generates the corresponding MHD modes. The bursts of magnetic activities
accompanied by accompanied by sudden runaway current drops observed in
experiments (see Figs.~\ref{119978_all} (a) and (b))
are probably related to the described phenomenon.    \par    

\section{Summary}
\label{summary}
Based on the analysis of numerous experimental data obtained in the TEXTOR
tokamak we have proposed a possible mechanism of the plasma disruption with
the formation of RE beams. The plasma disruption starts due to a 
large--scale magnetic stochasticity caused by nonlinearly excited of MHD modes
with low $(m,n)$ numbers ($m/n= 1/1, 2/1, 3/2$, $5/2, \dots$). The RE beam is
formed in the central plasma region confined by the intact magnetic
surface. Its location depends on the safety factor profile $q(\rho)$ and the
spectrum of MHD modes. In the cases of plasmas with the monotonic profile of
$q(\rho)$ and at the sufficiently small amplitude of the $m/n= 1/1$ mode  the
most stable RE beams are formed by the intact magnetic surface located between
the magnetic surface $q=1$ and the closest low--order rational 
surface $q=m/n>1$. Depending on the spectrum of magnetic perturbations
this rational magnetic surface could be one of these ones: $q=4/3$, $q=5/4$ or
$q=3/2$.  \par  

Such an outermost intact magnetic surface forms the transport barrier for
particles in the 
central plasma region. Electrons in this confined region are accelerated by
the inductive toroidal electric field. Such a situation occurs, for instance,
in plasma disruptions with runaway beams initiated by the Argon gas
injection. Heavy Ar atoms do not penetrate sufficiently deep into the
plasma and therefore they do not excite the $m/n=1/1$ mode with the amplitude
necessary to create the fully chaotic magnetic field. On the other hand the
injection of the lighter noble gases neon and helium does not generate
runaways since the light gases penetrate deeper into the plasma and excite the
large--amplitude ($m/n=1/1$) mode. \par 

During disruptions of tokamak plasmas with the reversed magnetic
shear RE beams can be formed in the central plasma region confined by the
shearless magnetic surface. The latter cannot be broken even at the relatively
large magnetic perturbations and it acts as a robust transport barrier to the
parallel motion of particles along chaotic magnetic field lines. One expects
that electrons confined by this intact magnetic surface form a relatively
stable RE beam with a large transversal size. Experimental observations
of RE beams with long confinement times during disruptions of plasma with the
reversed magnetic shear in the TFTR tokamak \citep{Fredrickson_etal_15}, 
probably, supports this expectation.  \par

Based on this scenario we proposed the models of the pre-disruption and
post--disruption plasmas with REs to study the processes of thermal and
currents quenches, the runaway current losses. The model of magnetic field was
proposed to describe the large--scale magnetic stochasticity due to
interaction of low--mode--number MHD modes. The radial transport of heat
and particles in a stochastic magnetic field are studied using the collisional
diffusional models. It was shown that the temperature drop during the fast
phase of disruption is caused by the radial heat transport determined by the
collisional electron transport in a stochastic magnetic field. We have
estimated a current decay time using the ambipolar collisional particle
transport model. The dynamics of RE orbits in a post--disruption plasma in the
presence of the inductive toroidal electric field is investigated by
integrating the equations of guiding center motion. We analyzed the effect of
the internal MHD mode and external RMPs on the topology of RE orbits.    \par 

The new model reproduces for the first time remarkably well the essential
features of the measurements:
\begin{itemize}
\item[{(a)}] The outer part of the plasma is clearly ergodized while the inner
  section is still intact. This agrees with the observation that the runaways
  are only seen in the inner half of the torus while they are obviously
  quickly lost from the outer part.  
\item[{(b)}] one observes a short tiny spike during the energy quench; we have
  described this spike previously; this spike is attributed to the loss of
  runaways born at the start up of the discharge from the ergodic zone. 
\item[{(c)}] In the case of disruptions caused by injection of light
  He and Ne gases the energy quench is due to processes directly related to
  the penetration of neutral impurity into the plasma; if heavier argon is
  injected fast electron transport in stochastic magnetic field is of more
  importance. Stochastic motion of plasma particles is also responsible for the
  current decay; however due to the ambipolarity of particle losses the ion
  motion essentially affects the duration of the CQ stage. The estimations of
  the energy quench and the current decay times based on the models presented
  agree well with observations. 
\item[{(d)}] The slow decay of the RE current in the
plateau phase is explained by the loss of runaways due to two effects: ($i$)
an outward shift of the runaways due to their continuous acceleration and the
subsequent loss at the wall; ($ii$) by the formation of a stochastic layer of 
high--energy REs at the beam edge in the presence of the $m/n=1/1$ MHD mode. 
\item[{(e)}] The effect of the external resonant magnetic perturbations on
  low-energy electrons (up to 5-10 MeV) is weak and does not cause their loss. 
\end{itemize}
%-----------------------------------------------------------------------
The new mechanism explains well the observed disruptions in present day
tokamaks. One can expect the following consequences, e.g., for ITER.
%-----------------------------------------------------------------------
\begin{itemize}
\item[{(1)}] The structure of the stochastic zone during the TQ
  allows persistence of preexisting runaways through this phase such that they
  act as seeds during the following phase of high loop voltage.
\item[{(2)}] The decay phase of the REs is rather long such that REs can
  acquire very high energy.
\item[{(3)}] External magnetic perturbations acting on REs seems are little
  promising unless the core of a RE beam can be ergodized. 
\item[{(4)}] A means for eliminating the REs completely is the injection of
  about $10^{25}$ molecules H$_2$ or D$_2$ into the discharge
  \citep{ITER:2007_3}. This massive gas injection may impose a heavy load
  on the cryo--pumping system. 
\end{itemize}
%-----------------------------------------------------------------------
 
\section*{Acknowledgments}
The authors gratefully acknowledge valuable discussions with W. Biel,
S. Brezinsek, O. Marchuk, Ph. Mertens, D. Reiser, D. Reiter,  A. Rogister, and
U. Samm. S.S.A. thanks to V. Igochine for consulting on the MHD mode structure
in tokamaks.  We also thank the anonymous referees for their constructive
criticisms, comments, and valuable suggestions.  

%-------------------------- References ---------------------------------
%\bibliographystyle{jpp}
%\bibliography{/home/ssa/Bibliography/references_plasmas}
%\end{document}
%-----------------------------------------------------------------------

\end{document}